\documentclass[aps,prc,twocolumn,showpacs,preprintnumbers,nofootinbib,float,superscriptaddress]{revtex4-1}
\usepackage{colortbl}
\usepackage{epsfig}
\usepackage[dvipsnames]{xcolor}
\usepackage{float,amsmath,amssymb}
\usepackage{graphicx}
\usepackage{url}
\usepackage{bbold}
\usepackage[utf8]{inputenc}
\usepackage{physics}
\usepackage{textcomp}
\usepackage[caption=false]{subfig}
\usepackage[labelfont=bf]{caption}
\usepackage{booktabs,multirow}

\DeclareCaptionLabelFormat{andtable}{#1~#2  \&  \tablename~\thetable}

\newcommand{\xpom}{x_{I\hspace{-0.3em}P}}

\newcommand{\gev}{\mathrm{GeV}}

\newcommand{\xbj}{{x}}

\newcommand{\rt}{{\mathbf{r}}}

\newcommand{\bt}{{\mathbf{b}}}

\newcommand{\xt}{{\mathbf{x}}}
\newcommand{\kt}{{\mathbf{k}}}
\newcommand{\yt}{{\mathbf{y}}}
\newcommand{\zt}{{\mathbf{z}}}

\newcommand{\jpsi}{$\mathrm{J}/\psi$ }
\newcommand{\jpsim}{\mathrm{J}/\psi}

\newcommand{\sa}[1]{\left\langle #1 \right\rangle}
\newcommand{\coh}{\gamma^* p\to \mathrm{J}/\psi p}
\newcommand{\incoh}{\gamma^* p\to \mathrm{J}/\psi p^*}
\newcommand{\appropto}{\mathrel{\vcenter{
  \offinterlineskip\halign{\hfil$##$\cr
    \propto\cr\noalign{\kern2pt}\sim\cr\noalign{\kern-2pt}}}}}
\DeclareMathOperator\arctanh{arctanh}
\DeclareMathOperator\Ei{Ei}
\newcommand{\cpm}[3]{{#1}_{-#2}^{+#3}}

\usepackage[breaklinks,colorlinks,citecolor=citcolor,urlcolor=blue,linkcolor=lcolor]{hyperref}
\definecolor{lcolor}{rgb}{0.5,0,0}
\definecolor{citcolor}{rgb}{0,0.3,0.0}
\usepackage[capitalise]{cleveref}
\usepackage{todonotes}

\begin{document}

\title{Exclusive \jpsi production off a dilute proton within a refined hotspot description}

\author{Heikki M\"antysaari}
\affiliation{Department of Physics, University of Jyv\"askyl\"a, P.O. Box 35, 40014 University of Jyväskylä, Finland}
\affiliation{Helsinki Institute of Physics, P.O. Box 64, 00014 University of Helsinki, Finland}

\author{Anh Dung Le}
\affiliation{Department of Physics, University of Jyv\"askyl\"a, P.O. Box 35, 40014 University of Jyväskylä, Finland}
\affiliation{Helsinki Institute of Physics, P.O. Box 64, 00014 University of Helsinki, Finland}

\begin{abstract}
    We revisit the calculation of exclusive \jpsi  production in electron-proton scattering within the QCD dipole model, employing a refined description of the proton in terms of gluonic hot spots in the dilute regime. In contrast to earlier studies, we incorporate key missing elements: the event-by-event fluctuations in both the number of hot spots and their local saturation scales, as well as the first relativistic corrections to the \jpsi wave function. 
    We perform a Bayesian analysis  to constrain the model parameters using HERA data.
    These enhancements significantly improve the agreement with HERA data. Saturation scale fluctuations are found to be more important than fluctuations in the number of hot spots.  
    Including the first relativistic correction is found to be important, especially in order to describe the observed power-law behavior in the incoherent cross section at large $t$.
\end{abstract}

\maketitle

\section{Introduction}
\label{sec:intro}

Recent advances in Quantum Chromodynamics (QCD) have driven substantial progress in unraveling the internal structure of hadronic states -- particularly protons and nuclei -- in terms of their fundamental constituents: quarks and gluons. A central goal of this line of research is to elucidate how hadronic substructure influences various high-energy scattering observables, thereby bridging theoretical QCD calculations with experimental measurements. Experimentally, insights into hadronic substructure are indispensable for interpreting a wide range of data from high-energy scattering experiments. 
These insights rely on data from the H1 and ZEUS experiments at the DESY HERA or from the active experimental program of ultraperipheral collisions at the LHC and RHIC~\cite{Baltz:2007kq,Klein:2020nvu}. In the future, the Electron-Ion Collider (EIC)~\cite{Accardi:2012qut} is expected to provide data which shall refine our understanding of the internal structure of protons and nuclei. 

The most basic description of a proton is the valence quark picture, in which the proton is treated as a bound state of three constituent quarks. This simplified model provides a reasonable approximation for the proton in low-energy processes (see, e.g., Refs.~\cite{Dumitru:2019qec,Dumitru:2020gla,Dumitru:2021tvw}). However, in the context of high-energy scattering -- such as deep inelastic scattering (DIS) or proton–proton collisions at current and future collider energies -- a more refined description becomes necessary. At high energies, the proton is better described as a highly evolved quantum state, where additional partons (gluons and sea quarks) are dynamically generated. These partonic cascades emerge from the original valence quarks and are described by perturbative QCD formalisms, such as the DGLAP (Dokshitzer–Gribov–Lipatov–Altarelli–Parisi)~\cite{Gribov:1972ri,Lipatov:1974qm,Altarelli:1977zs,Dokshitzer:1977sg} or  BFKL (Balitsky–Fadin–Kuraev–Lipatov)~\cite{Lipatov:1976zz,Kuraev:1976ge,Kuraev:1977fs,Balitsky:1978ic} evolutions.

A recent widely adopted and phenomenologically successful approach to modeling this evolved structure is the hotspot model, in which the proton is represented as an ensemble of localized regions of high gluon density, commonly referred to as gluonic hot spots \cite{Albacete:2016pmp,Albacete:2017ajt,Mantysaari:2016ykx,Mantysaari:2016jaz,Mantysaari:2017dwh, Cepila:2016uku,Cepila:2017nef,Mantysaari:2018zdd,Traini:2018hxd,Kumar:2021zbn,Kumar:2022aly,Demirci:2021kya,Demirci:2022wuy,Mantysaari:2022ffw}. In this picture, the gluon content is not uniformly distributed but is instead concentrated in spatially confined domains with a small transverse radius. This reflects the fluctuating nature of the proton’s internal structure at high energies and plays a critical role in describing observables sensitive to event-by-event fluctuations, such as diffractive  cross sections (see, e.g., Refs.~\cite{Mantysaari:2016ykx,Cepila:2016uku,Mantysaari:2018zdd}) and particle correlations in proton-proton and proton-lead collisions (see e.g.~\cite{Schenke:2014zha,Albacete:2017ajt,Mantysaari:2017cni,Moreland:2018gsh}). For a review of the proton (and nuclear) shape fluctuations, see  Ref.~\cite{Mantysaari:2020axf}.

Many contemporary descriptions of hot spots in the proton structure (see, e.g., Refs.~\cite{Mantysaari:2016ykx,Mantysaari:2018zdd,Demirci:2021kya,Demirci:2022wuy}) are grounded in the Color Glass Condensate (CGC) effective field theory~\cite{Iancu:2003xm}. The CGC framework provides a natural   theoretical foundation for analyzing various sources of fluctuations within hadronic wave functions. In this formalism, partons carrying large momentum fractions $x$ are treated as static color charges that radiate gluons at small-$x$. Due to the high occupancy of these small-$x$ gluons, they can be effectively described by classical gluon fields. To model the distribution of the color-charge sources, Gaussian correlations are assumed, as in the McLerran-Venugopalan (MV) model~\cite{McLerran:1993ka,McLerran:1993ni,McLerran:1994vd}. In the full nonlinear regime, hotspot models typically require numerical implementation due to the intrinsic complexity of the underlying dynamics. In the dilute limit where observables such as scattering cross sections can be expressed as expansions into the lowest non-trivial order in the gluon field strength, analytical computations become feasible and have been successfully carried out in Refs.~\cite{Demirci:2021kya,Demirci:2022wuy}. However, a numerical implementation of the dilute-limit calculation of exclusive \jpsi photoproduction in electron-proton scattering in Ref.~\cite{Demirci:2022wuy} has demonstrated poor agreement with the available HERA data. This discrepancy is unlikely to stem from the use of the dilute-limit approximation. In fact, some previous studies \cite{Bautista:2016xnp,Penttala:2024hvp} have indicated that the linearized framework can yield satisfactory fits for exclusive heavy vector meson production in the scattering off the proton target.  Instead, the observed mismatch is more plausibly attributed to the absence of certain components in the model, such as the fluctuations in the number of hot spots and in the local saturation scales. 

In the present study, we revisit the above-mentioned proton hotspot model in the dilute limit and its application to  exclusive \jpsi production in electron-proton scattering.  Exclusive vector meson production is known to be highly sensitive to fluctuations in the hadronic target wave function. As such, it serves as a powerful probe of the proton’s internal gluonic structure—especially in the incoherent channel. To improve upon previous implementations of the model, we introduce additional sources of  fluctuations: namely, event-by-event variations in both the number of gluonic hot spots and in the local saturation scales. These ingredients, which were absent in the earlier analysis presented in Ref.~\cite{Demirci:2022wuy}, are expected to play a crucial role in shaping the incoherent scattering spectrum.

Moreover, we examine the impact of relativistic corrections to the light-cone wave function of the \jpsi. Specifically, we incorporate the first-order relativistic correction to the fully non-relativistic approximation, with the aim of quantifying its effect on the predicted production cross sections. This correction addresses another missing component in the previous calculation and is anticipated to improve the accuracy of the model.

By construction, the model contains several free parameters that must be constrained in order to make quantitative predictions. In previous works~\cite{Demirci:2021kya,Demirci:2022wuy}, these parameters were manually chosen based on physical intuition and prior experience, rather than being systematically determined by comparing with the available experimental data. In the present analysis, we adopt a more rigorous and data-driven approach by employing Bayesian inference to extract the model parameters. Bayesian analysis has recently gained significant traction in the fields of high-energy QCD and nuclear physics, where it has been successfully applied to various problems (see, e.g., ~\cite{Mantysaari:2022ffw,Parkkila:2021yha,JETSCAPE:2020mzn,Casuga:2023dcf}). In a particularly relevant application, Ref.~\cite{Mantysaari:2022ffw} employed Bayesian inference to constrain the parameters of a hotspot model for exclusive \jpsi production in electron-proton scattering within the full nonlinear CGC framework. Their analysis yielded a set of parameters that provided excellent agreement with H1 photoproduction data at low momentum transfer~\cite{H1:2013okq}, demonstrating the power of Bayesian inference in extracting reliable, uncertainty-quantified model inputs from experimental measurements.    

The paper is organized as follows. In \cref{sec:general_formulation}, we provide an overview of the QCD dipole picture for exclusive \jpsi production in electron-proton scattering, which serves as a powerful theoretical framework for both photoproduction and electroproduction regimes. This section also introduces the modeling of the proton wave function in terms of gluonic hot spots, along with a detailed discussion of the projectile structure -- specifically, the overlaps between the virtual photon and  \jpsi light-cone wave functions. In \Cref{sec:excl_Jpsi_prod}, we derive the relevant expressions for the differential cross sections associated with both coherent and incoherent exclusive \jpsi productions. The Bayesian inference framework employed for the parameter estimation is outlined in \cref{sec:bayesian}, and the resulting posterior distributions, along with the discussions of the model's performance against the HERA data, are presented in \cref{sec:numerics}. Finally, \cref{sec:conclusion} summarizes our main findings and discusses potential relevant perspectives.      

\section{Exclusive $\mathrm{J}/\psi$ production
in the dipole picture and proton hot spots}  

\label{sec:general_formulation}

\subsection{Vector meson production}

\begin{figure}[ht!]
    \centering
    \includegraphics[width=0.99\linewidth]{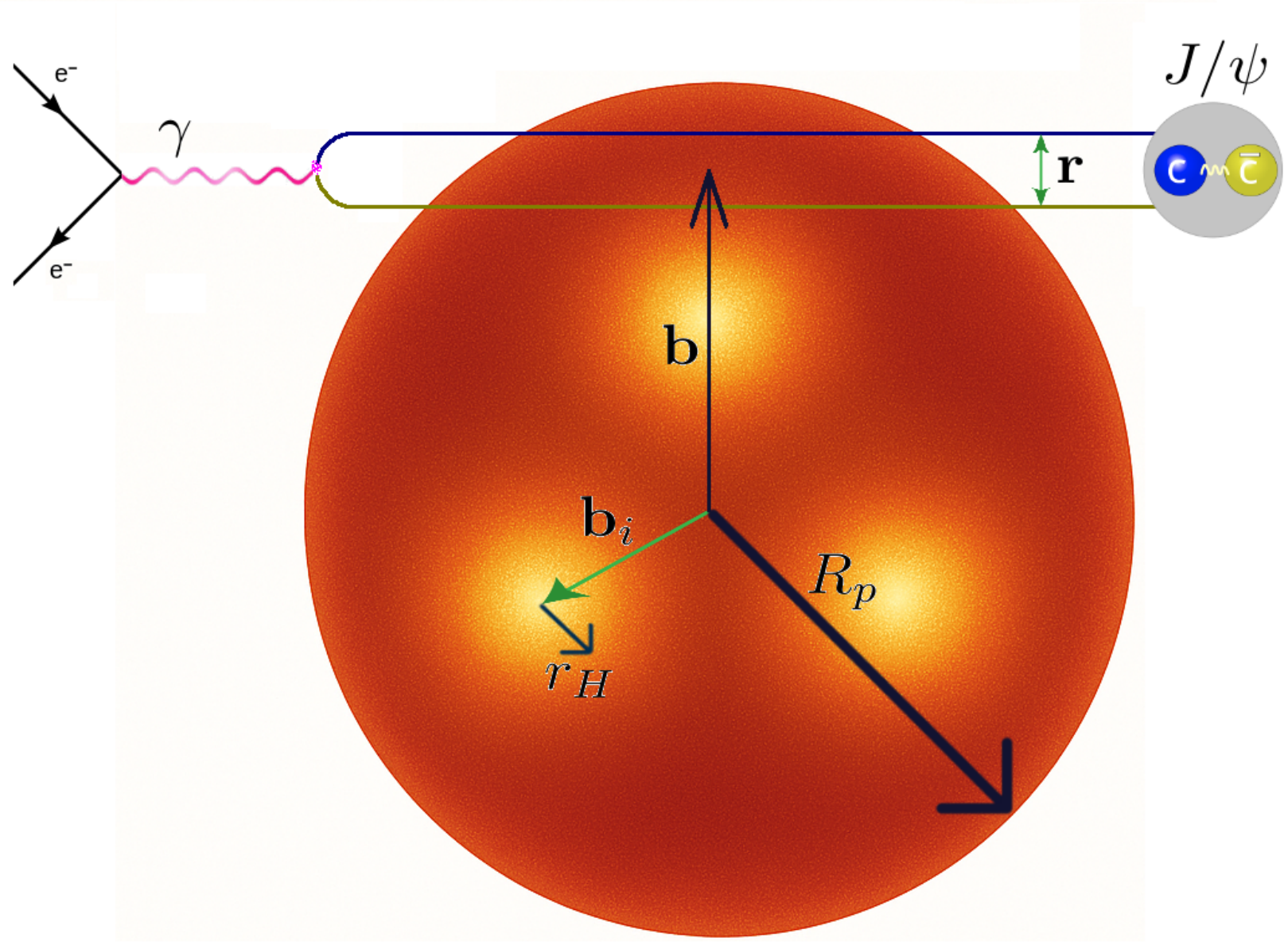}
    \caption{Illustration of exclusive \jpsi production in electron–proton scattering, modeled within the QCD dipole framework incorporating a gluonic hotspot structure of the target proton. The hotspots are depicted as yellow regions with characteristic size $r_H$. Additional geometric variables shown in the figure are described in the main text.}
    \label{fig:illustration}
\end{figure}

We consider  exclusive \jpsi production in electron-proton scattering. Within the single-photon exchange approximation, the interaction between the electron and the proton is mediated by a virtual (or quasi-real) photon $\gamma^{*}$ of virtuality $Q^2$. In the framework of the QCD dipole model, the virtual photon fluctuates into a $c\Bar{c}$ dipole of transverse size $\rt$, which then interacts with the proton target. At leading order, this dipole undergoes elastic scattering and subsequently hadronizes into a \jpsi meson in the final state (see \cref{fig:illustration}). The scattering amplitude depends on the polarization of the incoming photon -- either longitudinal (L) or transverse (T) -- and is given by~\cite{Kowalski:2006hc,Mantysaari:2020axf,Hatta:2017cte}
\begin{equation}
\label{eq:scattering_amplitude}
\begin{aligned}
    &\mathcal{A}_{L,T}^{\mathrm{J}/\psi}(W,Q^2,t) \\
    & = i \int \dd^2\bt \int \dd^2\rt \int \frac{\dd z}{4\pi}\ e^{-i\Delta\cdot\left[\bt + \left(\frac{1}{2}-z\right)\rt\right]} \\
    & \times \left(\Psi_{\gamma^*}^{*}\Psi_{\mathrm{J}/\psi}\right)_{L,T} (Q^2,\rt,z) ~ \frac{\dd^2\sigma_{\rm dip}}{\dd^2\bt} (\rt,\bt,\xpom). 
\end{aligned} 
\end{equation}
%
Here $z$ is the fraction of the large photon light-cone $``+"$  momentum  carried by the quark in the dipole, and $W$ is the center-of-mass energy of the $\gamma^*p$ process. The transverse momentum transfer $\Delta$ is the Fourier conjugate to $\bt + \left(\frac{1}{2}-z\right)\rt$ which is the center-of-mass of the dipole, when $\bt$ is the center of the $q\bar q$ system. Furthermore, in a high-energy scattering $t=-\Delta^2$. The target momentum fraction  $\xpom$ is related to the invariants $Q^2, W$ and $t$ as 
\begin{equation}
    \xpom \equiv \frac{Q^2+M_{\mathrm{J}/\psi}^2 - t}{Q^2+W^2 - m_p^2},
\end{equation} 
where $m_p$ is the proton mass.

According to the dipole factorization formalism in \cref{eq:scattering_amplitude}, the calculation of the scattering amplitude reduces to evaluating two key components: the overlap of the photon and vector meson wave functions, $\left(\Psi_{\gamma^*}^{*}\Psi_{\mathrm{J}/\psi}\right)$, and the dipole–proton cross section at a fixed impact parameter, $\frac{\dd^2\sigma_{\rm dip}}{\dd^2\bt}$. The wave-function overlap encapsulates the structure of the probing system -- namely, the transition of the virtual photon into a quark–antiquark dipole and its subsequent formation into a vector meson such as the $\mathrm{J}/\psi$ in the current study. Meanwhile, the dipole–proton cross section contains information about the internal structure of the target proton. In QCD, the proton is described as a highly complex quantum state composed of quarks and gluons, capable of exhibiting an infinite number of possible configurations. As a result, the microscopic configuration of the proton can fluctuate from one scattering event to another, leading to event-by-event variations in the interaction dynamics. 

The exclusive production of the \jpsi meson -- and more generally, of any meson -- in high-energy $e+p$ scattering can be categorized into two classes according to the final state of the target. If the target remains intact, the process is called coherent production. In this case, the scattering is quasi-elastic, and the corresponding cross section reflects an average over all possible configurations of the target without resolving its internal structure,
\begin{equation}
\label{eq:coherent_xsection}
    \frac{\dd \sigma^{\coh}_{L,T}}{\dd t} = \frac{1}{16\pi} \left|\sa{\mathcal{A}_{L,T}^{\mathrm{J}/\psi}(W,Q^2,t)}_{\rm target}\right|^2.
\end{equation}
Coherent production is therefore sensitive to the average spatial distribution of gluons within the target. In contrast, if the target breaks up after the interaction while still leaving a rapidity gap in the final state, the process is referred to as incoherent production. The corresponding cross section can be expressed in terms of the variance of the amplitude as
\begin{equation}
\label{eq:incoherent_xsection}
\begin{aligned}
    \frac{\dd \sigma^{\incoh}_{L,T}}{\dd t} & = \frac{1}{16\pi} \left[ 
    \sa{\left|\mathcal{A}_{L,T}^{\mathrm{J}/\psi}(W,Q^2,t)\right|^2}_{\rm target}  \right. \\
    & \left. \left. - 
    \left|\sa{\mathcal{A}_{L,T}^{\mathrm{J}/\psi}(W,Q^2,t)}_{\rm target}\right|^2 \right. \right].
\end{aligned}
\end{equation}
This production channel is then sensitive to fluctuations in the target’s substructure and provides valuable insight into event-by-event variations in its internal gluonic configuration. See Refs.~\cite{Good:1960ba,Miettinen:1978jb,Caldwell:2010zza,Mantysaari:2016ykx,Mantysaari:2020axf,Klein:2023zlf} for more detailed discussions.

\subsection{A hot spot description of the proton}

In the CGC approach, the wave function of an evolved proton involves large-$\xbj$ color sources and small-$\xbj$ gluon fields generated by the former. We can assume that the color charges are located around $N_h$ centers $\bt_i$ ($i=1,\hdots,N_h$) measured from the center of mass of the proton, which can be referred to as the hot spots. The transverse-space color charge density $\mu(\xt)$ within a hot spot is taken to be Gaussian,
\begin{equation}
 \label{eq:hotspot_profile}
    \begin{aligned}
        \mu^2(\xt) = \frac{\mu_0^2 \mathcal{Q}_s^2 e^{-\sigma_s^2/2} }{2\pi r_H^2} \exp\left(-\frac{\xt^2}{2r_H^2}\right),
    \end{aligned}
\end{equation}
where $r_H$ is the hot spot radius and $\mu_0^2$ is a parameter characterizing the average amount of color charge in the hot spot (average saturation scale $\sa{Q_s}$). 

The factor $\mathcal{Q}_s^2=Q_s^2/\sa{Q_s^2}$ 
describes the saturation scale fluctuations.
Beyond the calculation in Refs.~\cite{Demirci:2021kya,Demirci:2022wuy}, we include this source of fluctuations in the current analysis. The event-by-event $Q_s^2$ fluctuations are taken to follow the log-normal distribution of width $\sigma_s$ as in Refs.~\cite{McLerran:2015qxa,Bzdak:2015eii,Mantysaari:2016jaz}:
\begin{equation}
\label{eq:Qs_fluct}
    P\left[\log( \mathcal{Q}_s^2) \right] = \frac{1}{\sigma_s\sqrt{2\pi}} \exp\left[ -\frac{\log^2( \mathcal{Q}_s^2)}{2\sigma_s^2}\right].
\end{equation}
The expectation value of $\mathcal{Q}_s^2$ computed from the log-normal distribution given in \cref{eq:Qs_fluct} is $e^{\sigma_s^2/2}$, rather than $1$. To correct for this effect and ensure that the average $\mathcal{Q}_s^2$ remains unchanged, we scale the values of $\mathcal{Q}_s^2$ generated from this distribution by the inverse of this factor. This normalization procedure accounts for the appearance of the compensating factor $e^{-\sigma_s^2/2}$ in \cref{eq:hotspot_profile}.

The color charge density $\rho^a(\xt)$ at a transverse position $\xt$ fluctuates on an event-by-event basis. Based on the MV model \cite{McLerran:1993ka,McLerran:1993ni,McLerran:1994vd}, we assume Gaussian statistics for the distribution of the color charges, according to which the one-point and two-point correlators of color charge density read
\begin{subequations}
\label{eq:color_charge_fluct}
    \begin{equation}
    \label{eq:cgc_correlatior_1p}
        \sa{\rho^a(\xt)}_\mathrm{CGC} = 0,    
    \end{equation}
    \begin{equation}
    \label{eq:cgc_correlatior_2p}
    \begin{aligned}
        \sa{\rho^a(\xt)\rho^b(\yt)}_\mathrm{CGC} &= \frac{3}{N_h}\sum_{i=1}^{N_h} \mu^2\left(\frac{\xt+\yt}{2} - \bt_i\right) \\ 
        &\qquad\qquad\times\delta^{(2)}(\xt-\yt)\delta^{ab}.
    \end{aligned}    
    \end{equation}
\end{subequations} 
Note that here the longitudinal dependence is implicitly integrated out, and the average over the $Q_s$ fluctuation has not yet been performed in these correlators. The two-point correlator (\ref{eq:cgc_correlatior_2p}) is normalized to the number of hot spots $N_h$ to avoid the increase of color charge in the proton as the number of hot spots increases. The factor $3$ in \cref{eq:cgc_correlatior_2p} is to recover the unnormalized correlator in Ref.~\cite{Demirci:2021kya,Demirci:2022wuy} in which the number of hot spots was fixed at the value $N_h=3$. The condition (\ref{eq:cgc_correlatior_1p}) comes from the requirement of color neutrality. The Dirac delta function in \cref{eq:cgc_correlatior_2p} reflects a feature of the MV model in which two color charges at two different transverse positions are not correlated. Higher-order correlators can then be calculated from \cref{eq:color_charge_fluct} using Wick's theorem.  

The transverse positions $\bt_i$ ($i=1,\hdots,N_h$) of the hot spots also fluctuate on an event-by-event basis according to a Gaussian distribution given by
\begin{equation}
\label{eq:hotspot_distribution}
    T_p(\bt_i) = \frac{1}{2\pi R_p^2} \exp\left(-\frac{\bt_i^2}{2R_p^2}\right),
\end{equation}
where $R_p$ is the effective size of the proton. Following \cite{Demirci:2021kya,Demirci:2022wuy}, we require the center of mass of the hot spots to coincide with the center of mass of the proton, $({1}/{N_h})\sum_{i=1}^{N_h} \bt_i = 0$. 

In addition to the saturation scale fluctuation, we also allow the number of hot spots $N_h$ to vary from event to event around an expected value $\overline{N}_h >1$. This type of fluctuation is modeled using a zero-truncated Poisson distribution, following the approach described in Ref.~\cite{Cepila:2017nef}: the probability of having $N_h$ hot spots is
\begin{equation}
\label{eq:zero_truncated_Poisson}
    P_\Lambda(N_h) = \frac{\Lambda^{N_h}}{(e^{\Lambda}-1)N_h!}.
\end{equation}
Here $\Lambda > 0$ solves $\Lambda/(1-e^{-\Lambda}) = \overline{N}_h$. 

With the above description of the proton wave function, the average over target configurations in \cref{eq:coherent_xsection,eq:incoherent_xsection} reads
\begin{equation}
\label{eq:double_average}
\begin{aligned}
    & \sa{\mathcal{O}}_{\rm target} = \sum_{N_h=1}^{+\infty} P_{\Lambda}\left(N_h\right) \left(\frac{2\pi R_p^2}{N_h}\right)\\ 
    &\times \int \left[\prod_{k=1}^{N_h} \dd^2\bt_k T_h(\bt_k)\right] \delta^{(2)}\left(\frac{1}{N_h}\sum_{i=1}^{N_h}\bt_i\right) \\
    & \times \int \left\{\prod_{j=1}^{N_h}\dd\left[\log(\mathcal{Q}_{s,j}^2)\right] P\left[\log( \mathcal{Q}_{s,j}^2) \right] \right\} \sa{\mathcal{O}}_{CGC},
\end{aligned}
\end{equation}
where the factor $(2\pi R_p^2)/N_h$ is to ensure the normalization condition $\sa{\mathbf{1}}_{\rm target} = 1$.

To describe the data across various values of $Q^2$ and $W$, the model should account for the energy evolution of the system. This can be achieved by incorporating the JIMWLK~\cite{Jalilian-Marian:1996mkd,Jalilian-Marian:1997jhx,Jalilian-Marian:1997qno,Iancu:2000hn,Ferreiro:2001qy,Iancu:2001ad,Iancu:2001md} or BK~\cite{Balitsky:1995ub,Kovchegov:1999yj} evolutions into the framework, or by appropriately modeling the energy dependence of the relevant parameters (see, e.g., \cite{Cepila:2017nef,Cepila:2016uku,Kumar:2022aly}). However, in this work, we focus specifically on the effects of fluctuations in the saturation scale $Q_s$ and the number of hot spots $N_h$, as well as of the first relativistic correction to the exclusive \jpsi production within the dilute proton hotspot model. To this end, we fit the model to data at fixed values of $Q^2$ and $W$, allowing the momentum transfer $t$ to vary within a limited range. In this kinematics, the energy dependence can be considered negligible.  

\subsection{Wave function overlaps}

The calculation of the wave function overlaps that appear in \cref{eq:scattering_amplitude} requires information on the $c\Bar{c}$ wave functions of the virtual photon $\gamma^*$ and of $\mathrm{J}/\psi$. The branching $\gamma^* \to c\bar c$ is  a simple perturbative QED process and the photon wave function can be computed  using the light-cone perturbation theory~\cite{Brodsky:1997de,Kovchegov_Levin_2012,Angelopoulou:2023qdm}. They are quoted in~\cref{sec:appendix_gamma_wf}.

The $\mathrm{J}/\psi$ light cone wave function is a non-perturbative object. As $c$ is a heavy quark, the relative velocity of the $c\bar c$ pair is parametrically small. 
Thus, one can expand the $\mathrm{J}/\psi$ wave function in terms of this small parameter (the so-called non-relativistic QCD (NRQCD) expansion). In Ref.~\cite{Lappi:2020ufv}, the authors proposed a parameterization for the heavy vector meson wave functions based on NRQCD long-distance matrix elements. Using this method, they extracted the $q\bar q$ component of the wave functions up to the first relativistic correction. This correction was shown to have a significant effect on $\mathrm{J}/\psi$ production phenomenology. On the other hand, in a recent analytical hotspot model calculation for exclusive $\mathrm{J}/\psi$ production~\cite{Demirci:2022wuy}, the authors used the fully non-relativistic limit of the wave function. For highly massive states like $b\bar b$, this is a very good approximation. However, for charmonium states like $\mathrm{J}/\psi$, the relativistic correction may have a significant effect. Therefore, we shall include the first relativistic correction in our calculation to assess its effect on the production cross sections. The corresponding wave functions are quoted in~\cref{sec:appendix_Jpsi_wf}.           
The wave function overlaps can be computed using the wave functions quoted in~\cref{sec:appendix_gamma_wf,sec:appendix_Jpsi_wf}. The results after integrating over the quark's momentum fraction $z$ read~\cite{Lappi:2020ufv}

\begin{subequations}
\label{eq:wf_overlap}
    \begin{equation}
    \label{eq:wf_overlap_L}
    \begin{aligned}
        & \int_0^1 \frac{\dd z}{4\pi} \left(\Psi_{\mathrm{J}/\psi}^*\Psi_{\gamma^*}\right)_L^{c\bar c} e^{i\left(z-\frac{1}{2}\right)\rt\cdot\Delta} = -\frac{e_c e Q\sqrt{2N_c}}{8\pi\sqrt{m_c}} \\
        & \times \left\{ AK_0(\varepsilon |\rt|) + \frac{B}{m_c^2}\left[ \frac{9}{2}K_0(\varepsilon |\rt|) + m_c^2|\rt|^2 K_0(\varepsilon |\rt|)  \right.\right. \\
        &\left.\left. - \frac{Q^2|\rt|}{4\varepsilon}K_1(\varepsilon|\rt|) + \frac{1+\cos2\theta_r}{8}(|\rt|^2\Delta^2)K_0(\varepsilon|\rt|)\right]\right\},
    \end{aligned}
    \end{equation}
    \begin{equation}
    \label{eq:wf_overlap_T}
    \begin{aligned}
        &\int_0^1 \frac{\dd z}{4\pi} \left(\Psi_{\mathrm{J}/\psi}^*\Psi_{\gamma^*}\right)_T^{c\bar c} e^{i\left(z-\frac{1}{2}\right)\rt\cdot\Delta} = -\frac{e_c e \sqrt{2m_cN_c}}{4\pi} \\
        &\times \left\{ AK_0(\varepsilon |\rt|) +\frac{B}{m_c^2} \left[\frac{7}{2} K_0(\varepsilon |\rt|) + m_c^2|\rt|^2K_0(\varepsilon |\rt|) \right.\right. \\
        & \left.\left.- \frac{|\rt|}{2\varepsilon} \left(Q^2 + 2m_c^2\right)K_1(\varepsilon |\rt|) + \frac{1+\cos2\theta_r}{8}|\rt|^2\Delta^2K_0(\varepsilon|\rt|)\right] \right\},
    \end{aligned}
    \end{equation}
\end{subequations}
where $N_c$ is the number of colors, $e_c = 2/3$ is the fractional charge of the charm quark, $\theta_r$ is the angle between $\rt$ and $\Delta$, and $\varepsilon^2 = \frac{Q^2}{4} + m_c^2$. The coefficients $A$ and $B$ read
\begin{equation}
\label{eq:AB_values}
    \begin{aligned}
        A \approx 0.213 ~\gev^{3/2},~
        B \approx -0.0157 ~\gev^{7/2}.
    \end{aligned}
\end{equation}
In the fully non-relativistic limit, $B$ vanishes and the corresponding value of A is slightly different from the above value, $A_{NR}\approx0.211 ~\gev^{3/2}$~\cite{Demirci:2022wuy}. At $|\rt| = r_m\sim 0.6 \rm ~fm$, the wave function overlaps (\ref{eq:wf_overlap}) become negative when the first relativistic correction is taken into account. We will set the overlaps to zero at $|\rt|>r_m$ to avoid the appearance of the node in the $\mathrm{J}/\psi$ wave functions. In Ref.~\cite{Lappi:2020ufv}, this cut-off was checked to have a negligible effect when convoluting with the IPsat dipole cross-section. In this study, the effect of this cut-off is more significant (although still not large), as the dipole cross-section derived in the dilute limit in the next section does not saturate at large $|\rt|$. 

\section{Exclusive scattering in the dilute limit}
\label{sec:excl_Jpsi_prod}

\subsection{Coherent production}

The dipole-proton cross section at a fixed impact parameter can be expressed in terms of Wilson lines $V(\xt)$ as
\begin{equation}
\label{eq:dipole_xsection_wilson}
\begin{aligned}
    \frac{\dd^2\sigma_{\rm dip}}{\dd^2\bt} (\rt,\bt) = 2\left(1 - \frac{1}{N_c}\Tr{V(\xt)V^{\dagger}(\yt)}\right),
\end{aligned}
\end{equation}
where the impact parameter $\bt$ and the transverse dipole size $\rt$ are related to the transverse coordinates $\xt$ and $\yt$ of the (anti)quark as
\begin{equation}
    \bt = \frac{\xt + \yt}{2}, ~ \rt =\xt - \yt.
\end{equation}
The light-like fundamental Wilson line $V(\xt)$ represents the eikonal propagation of a quark at transverse position $\xt$ through the target color field, and is defined as a path-ordered exponential along the plus direction,
\begin{equation}
\label{eq:Wilson_line}
    \begin{aligned}
        V(\xt) &= \mathcal{P}_{+} \exp\left[ig \int_{-\infty}^{+\infty} \dd z^{+} A_{a}^{-}(z^+,\xt)t^a \right] \\*
        &= \mathcal{P}_{+} \exp\left[ig \int \dd[2]{\zt} G(\xt-\zt) \rho_{a}(\zt)t^a \right],
    \end{aligned}
\end{equation}
where $g$ is the strong coupling constant and the Green's function is given by
\begin{equation}
\label{eq:Green_function}
    G(\xt - \zt) = \int \frac{\dd^2\kt}{(2\pi)^2} \frac{e^{i\kt\cdot(\xt - \zt)}}{\kt^2 + m^2},
\end{equation}
with an infrared cut-off $m^2$ to suppress the long-range Coulomb tail of the color field. Following Ref.~\cite{Demirci:2022wuy}, the CGC average of the dipole cross section in the dilute limit employing~\cref{eq:color_charge_fluct} reads
\begin{equation}
\label{eq:cgc_average_dipole_xsec}
    \begin{aligned}
        &\sa{ \frac{\dd^2\sigma_{\rm dip}}{\dd^2\bt}}_{\rm CGC} = \frac{3g^2C_F}{N_h}e^{-\sigma_s^2/{2}}\int \dd^2\zt \\ &\times \sum_{i=1}^{N_h}\mu^2(\zt-\bt_i)
         \left[G\left(\bt+\frac{\rt}{2}-\zt\right)G\left(\bt+\frac{\rt}{2}-\zt\right) \right.\\
        &\left. +~ G\left(\bt-\frac{\rt}{2}-\zt\right)G\left(\bt-\frac{\rt}{2}-\zt\right) \right. \\
        &\left. -~ 2G\left(\bt+\frac{\rt}{2}-\zt\right)G\left(\bt-\frac{\rt}{2}-\zt\right) \right],
    \end{aligned}
\end{equation}
where $C_F = (N_c^2-1)/(2N_c)$ is the fundamental Casimir of $su(N_c)$. The additional factor of $e^{-\sigma_s^2/{2}}$ enforces the same average $Q_s^2$ independently of the $Q_s$ fluctuations, see the discussion in \cref{sec:excl_Jpsi_prod}.
This factor cancels out exactly when averaging over $Q_s^2$ fluctuations using \cref{eq:Qs_fluct}.
Therefore, the $Q_s^2$ fluctuations have no effect on the average dipole cross section in the dilute limit. 

Performing the average over the hot spot transverse positions as in Ref.~\cite{Demirci:2022wuy}, we obtain
\begin{equation}
\label{eq:target_average_dipole_xsec}
\begin{aligned}
    &\sa{ \frac{\dd[2]{\sigma_{\rm dip}}}{\dd^2\bt}}_{\rm target} =  ~3g^2C_F \sum_{N_h=1}^{+\infty} P_{\Lambda}\left(N_h\right) \\
    &\times \int \dd[2]{\zt}  F_1(\zt)\Omega\left(\bt+\frac{\rt}{2}, \bt-\frac{\rt}{2},\zt,\zt \right),
\end{aligned}
\end{equation}
where
\begin{equation}
\label{eq:Omega_function}
    \begin{aligned}
        &\Omega\left(\xt, \yt, \zt_1,\zt_2 \right) \equiv G(\xt - \zt_1)G(\xt - \zt_2) \\
        &+ G(\yt - \zt_1)G(\yt - \zt_2) - 2G(\xt - \zt_1)G(\yt - \zt_2) 
    \end{aligned}
\end{equation}
and
\begin{equation}
\label{eq:Omega_function}
    \begin{aligned}
        &F_1\left(\zt \right) \equiv  \frac{\mu_0^2}{2\pi\left(r_H^2 + \frac{N_h-1}{N_h}R_p^2\right)}\exp\left[-\frac{1}{2}\frac{\zt^2}{r_H^2 + \frac{N_h-1}{N_h}R_p^2}\right].
    \end{aligned}
\end{equation}
The function $F_1(z)$ is the one-point function of the color charge  content in the hot spots averaging over their spatial distribution, $F_1(z) = \sum_{i=1}^{N_h}\sa{\mu^2(z-\bt_i)}_{hs}$. After the Fourier transform, we obtain
\begin{equation}
\label{eq:dipole_xsec_fourier_transform}
    \begin{aligned}
        &\int \dd[2]{\bt} e^{-i\bt\cdot\Delta} \sa{ \frac{\dd^2\sigma_{\rm dip}}{\dd^2\bt}}_{\rm target} \\
        = &\frac{3g^2\mu_0^2C_F}{\pi} \exp\left[-\frac{1}{2}\left(r_H^2+R_p^2\right)\Delta^2\right] \\
        & \times \left[\cos\left(\frac{1}{2}\rt\cdot\Delta\right)\psi(0,\Delta) - \psi(\rt,\Delta)\right] \\
        & \times \sum_{N_h=1}^{+\infty} P_{\Lambda}(N_h) \exp\left(\frac{R_p^2\Delta^2}{2N_h}\right).
    \end{aligned}
\end{equation}
where
\begin{subequations}
    \begin{equation}
    \label{eq:psi_r}
    \begin{aligned}
        \psi(\rt,\Delta) & = \int_0^{1/2} \dd\eta \cos\left(\eta\rt\cdot\Delta\right) \\
        &\times \frac{|\rt|K_1\left(|\rt|\sqrt{-\Delta^2\eta^2 + \frac{\Delta^2}{4} + m^2}\right)}{\sqrt{-\Delta^2\eta^2 + \frac{\Delta^2}{4} + m^2}},
    \end{aligned}
    \end{equation}
    \begin{equation}
    \label{eq:psi_0}    
        \begin{aligned}
            \psi(0,\Delta) & = \int_0^{1/2} \frac{\dd\eta}{-\Delta^2\eta^2 + \frac{\Delta^2}{4} + m^2} \\
            & = \frac{2\arctanh{\frac{|\Delta|}{\sqrt{\Delta^2 + 4m^2}}}}{|\Delta|\sqrt{\Delta^2 + 4m^2}}.
        \end{aligned}
    \end{equation}
\end{subequations}
The coherent exclusive \jpsi cross section  can now be  computed from \cref{eq:scattering_amplitude,eq:coherent_xsection,eq:wf_overlap,eq:dipole_xsec_fourier_transform}. It reads
\begin{equation}
\label{eq:coherent_xsec_final}
    \begin{aligned}
        &\frac{\dd \sigma^{\gamma^*p\to \mathrm{J}/\psi p}_{T,L}}{\dd t} = \frac{9C_{T,L}^2}{16\pi}\frac{g^4\mu_0^4C_F^2}{\pi^2} e^{-\left(r_H^2 + R_p^2\right)|t|} \\
        &\qquad\qquad\times \left[\sum_{N_h=1}^{+\infty} P_{\Lambda}(N_h) e^{\frac{R_p^2}{2N_h}|t|}\right]^2 \mathcal{M}_{T,L}^2. 
    \end{aligned}
\end{equation}
where

    \begin{equation}
    \label{eq:coherent_C_LT}
    \begin{aligned}
        C_T = \frac{e_c e \sqrt{2m_cN_c}}{4\pi},~
         C_L = \frac{e_c eQ }{8\pi}\sqrt{\frac{2N_c}{m_c}},
    \end{aligned}
    \end{equation}    
    \begin{equation}
    \begin{aligned}
    \label{eq:coherent_MT}
        &\mathcal{M}_T = \int_{|\rt|<r_m} \dd^2\rt \left[\cos\left(\frac{1}{2}\rt\cdot\Delta\right)\psi(0,\Delta) - \psi(\rt,\Delta)\right] \\
        &\times \left\{ AK_0(\varepsilon |\rt|) +\frac{B}{m_c^2} \left[\frac{7}{2} K_0(\varepsilon |\rt|)  \right.\right. \\
        & \left.\left. + m_c^2|\rt|^2K_0(\varepsilon |\rt|) -  \left(\frac{Q^2}{4\varepsilon} + \varepsilon\right)|\rt|K_1(\varepsilon |\rt|) \right.\right.\\
        &\left.\left. + \frac{1+\cos2\theta_r}{8}(|\rt|^2\Delta^2)K_0(\varepsilon|\rt|)\right] \right\}, \\
    \end{aligned}
    \end{equation}   
    and
    \begin{equation}
    \label{eq:coherent_ML}
     \begin{aligned}
         \mathcal{M}_L &= \int_{|\rt|<r_m} \dd^2\rt \left[\cos\left(\frac{1}{2}\rt\cdot\Delta\right)\psi(0,\Delta) - \psi(\rt,\Delta)\right] \\
        &\quad\times \left\{ AK_0(\varepsilon |\rt|) +\frac{B}{m_c^2} \left[\frac{9}{2} K_0(\varepsilon |\rt|)  \right.\right. \\
        &\quad \left.\left. + m_c^2|\rt|^2K_0(\varepsilon |\rt|) - \frac{Q^2|\rt|}{4\varepsilon} K_1(\varepsilon |\rt|) \right.\right. \\
        &\quad \left.\left.+ \frac{1+\cos2\theta_r}{8}(|\rt|^2\Delta^2)K_0(\varepsilon|\rt|)\right] \right\}.
    \end{aligned}
\end{equation}
As was mentioned in the previous section, the cut-off ${|\rt|<r_m}$ in the above integrals is to avoid the large-$|\rt|$ region where the wave-function overlaps change sign (there should be no node in the \jpsi wave function). Comparing to the corresponding expression for the coherent cross sections in Ref.~\cite{Demirci:2022wuy}, we see that \cref{eq:coherent_xsec_final} involves additionally the dependence on the $N_h$ fluctuation and on the relativistic correction to the \jpsi non-relativistic wave function. The $Q_s$ fluctuations have no influence on the coherent diffractive cross sections (\ref{eq:coherent_xsec_final}) in the dilute limit. The dependence on the number of hot spot fluctuations factorizes as a squared summation which modifies the $t$ spectra and as such effectively rescales the proton's size. To see that more explicitly, let us perform a small-$|t|$ estimation. In this limit, we have
\begin{equation}
\begin{aligned}
    &\sum_{N_h=1}^{+\infty} P_{\Lambda}(N_h) e^{\frac{R_p^2}{2N_h}|t|} \approx \sum_{N_h=1}^{+\infty} P_{\Lambda}(N_h) \left(1+\frac{R_p^2|t|}{2N_h}\right)\\
    & = 1 + \frac{R_p^2|t|}{2}\mathcal{F}(\Lambda) \approx e^{\frac{R_p^2|t|}{2}\mathcal{F}(\Lambda)},
\end{aligned}
\end{equation}
where
\begin{equation}
    \label{eq:Nh_summation}
    \begin{aligned}
        \mathcal{F}(\Lambda) \equiv \sum_{N_h=1}^{\infty}P_{\Lambda}(N_h)\frac{1}{N_h}  = \frac{\Ei(\Lambda)-\log(\Lambda) -\gamma_E}{e^\Lambda - 1}, 
    \end{aligned}
\end{equation}
with $\Ei(x)$ being the exponential integral and $\gamma_E \approx 0,5772\hdots$ the Euler-Mascheroni constant. At small $|t|$, the differential cross section behaves as
\begin{equation}
\label{eq:coh_smallt_fluctnh}
    \frac{\dd \sigma^{\gamma^*p\to \mathrm{J}/\psi p}_{T,L}}{\dd t} \appropto e^{-\left[r_H^2 + R_p^2\left(1- \mathcal{F} (\Lambda)\right)\right]|t|}.
\end{equation}
In the case of no $N_h$ fluctuation, we have
\begin{equation}
\label{eq:coh_smallt_nofluctnh}
    \frac{\dd \sigma^{\gamma^*p\to \mathrm{J}/\psi p}_{T,L}}{\dd t} \propto e^{-\left[r_H^2 + R_p^2\left(1- \frac{1}{N_h}\right)\right]|t|}.
\end{equation}

The effect of fluctuations in the number of hot spots on the $t$-spectrum can be estimated as follows. Taking a typical value $\overline{N_h}=3$ corresponding to 
$\Lambda\approx2.82$, we get $1-\mathcal{F}(\Lambda)\approx 0.54< 1-1/3\approx 0.67$. This effectively makes the spectrum with $N_h$ fluctuations less steeply falling corresponding to a smaller proton compared to the case with a fixed number of $N_h=3$ hot spots. The factors $\mathcal{F}(\Lambda)$ and $1/N_h$ in \cref{eq:coh_smallt_fluctnh,eq:coh_smallt_nofluctnh} are a consequence of forcing the center of mass of the hot spots to coincide with the center of mass of the proton. In the limit of a large number of hot spots, or when the mentioned center-of-mass constraint is relaxed, these terms modifying the $t$ slope and the proton size would vanish. In the latter case, the $N_h$ fluctuations would have no effect on the coherent \jpsi  production in the dilute limit. From \cref{eq:coherent_xsec_final}, one can observe that this effect holds for the whole $t$ range. 
\subsection{Incoherent production}
In order to evaluate the incoherent cross section,~\cref{eq:incoherent_xsection},  we first need to calculate the following quantity:
\begin{equation}
\label{eq:squared_dipole_amplitude_expression}
    \sa{\frac{\dd^2\sigma_{\rm dip}}{\dd^2\bt}(\rt,\bt)\frac{\dd^2\sigma_{\rm dip}}{\dd^2\bt'}(\rt',\bt')}_{\rm target}.
\end{equation}
Expanding the squared amplitude to the order $\rho^4$ and performing the CGC average as in Ref.~\cite{Demirci:2022wuy}, we get
\begin{equation}
\label{eq:cgc_average_squared_amplitude}
    \begin{aligned}
       & \sa{\frac{\dd^2\sigma_{\rm dip}}{\dd^2\bt}(\rt,\bt)\frac{\dd^2\sigma_{\rm dip}}{\dd^2\bt'}(\rt',\bt')}_{\rm CGC} \\
        = & 9g^4\int \dd[2]{\zt_1} \dd[2]{\zt_2} \Bigg[C_F^2\Omega(\xt,\yt,\zt_1,\zt_1)\Omega(\xt',\yt',\zt_2,\zt_2)  \\
        &\left. + \frac{C_F}{2N_c} \Omega(\xt,\yt,\zt_1,\zt_2)\Omega(\xt',\yt',\zt_1,\zt_2) \right. \\
        & + \frac{C_F}{2N_c} \Omega(\xt,\yt,\zt_1,\zt_2)\Omega(\xt',\yt',\zt_2,\zt_1)\Bigg] \\
        &\times \left[\frac{1}{N_h^2}\sum_{i=1}^{N_h}\mu^2(\zt_1-\bt_i)\mu^2(\zt_2-\bt_i) \right. \\
        &\left. + \frac{1}{N_h^2}\sum_{i\neq j}^{N_h}\mu^2(\zt_1-\bt_i)\mu^2(\zt_2-\bt_j) \right].
    \end{aligned}
\end{equation}
Here $\rt'=\xt'-\yt'$ and $\bt'=(\xt'+\yt')/2$.
The first term proportional to $C_F^2$ in the first square bracket of \cref{eq:cgc_average_squared_amplitude} represents the contribution in which the color structures of the amplitude and of the conjugate amplitude are disconnected (DC). Meanwhile, the two remaining terms are connected (C) color structures. 

In the second bracket, the first term corresponds to the case where the color fields originate from the same hot spot. Averaging that contribution  over the hot spot transverse positions (denoted by $\langle \mathcal{O}\rangle_\mathrm{hs})$ and the $Q_s$ fluctuations ($\langle \mathcal{O}\rangle_{Q_s}$) gives
\begin{equation}
    \label{eq:F2_function}
    \begin{aligned}
        &F_2(\zt_1,\zt_2) \equiv \sum_{i=1}^{N_h} \sa{\mu^2(\zt_1-\bt_i)\mu^2(\zt_2-\bt_i)}_{\mathrm{hs},Q_s} \\
        &=  \sum_{i=1}^{N_h} e^{-\sigma_s^2}\int \dd\left[\log(\mathcal{Q}_{s,i}^2)\right]P[\log(\mathcal{Q}_{s,i}^2)] \mathcal{Q}_{s,i}^4 \\
        & \times \left(\frac{\mu_0^4 R_p^2N_h}{2\pi r_H^4 }\right) \int \prod_{k=1}^{N_h}\dd^2\bt_k T_h(\bt_k) \delta^{(2)}\left(\sum_{j=1}^{N_h}\bt_j\right) \\
        & \times \exp\left[-\frac{(\zt_1 - \bt_i)^2}{2r_H^2}-\frac{(\zt_2 - \bt_i)^2}{2r_H^2}\right] \\
        & = e^{\sigma_s^2}\left(\frac{\mu_0^4 R_p^2N_h^2}{2\pi r_H^4 }\right) \int \prod_{k=1}^{N_h}\dd^2\bt_k T_h(\bt_k)  \\
        &\times \exp\left[-\frac{(\zt_1 - \bt_1)^2}{2r_H^2}-\frac{(\zt_2 - \bt_1)^2}{2r_H^2}\right] \delta^{(2)}\left(\sum_{j=1}^{N_h}\bt_j\right) \\
        &= e^{\sigma_s^2} \left(\frac{\mu_0^2}{2\pi r_H^2 }\right)^2 \frac{N_h}{1 + \frac{N_h - 1}{N_h}\frac{2R_p^2}{r_H^2}} \\
        &\times \exp\left[-\frac{(\zt_1 + \zt_2)^2}{4r_H^2\left(1 + 2\frac{N_h-1}{N_h}\frac{R_p^2}{r_h^2}\right)} - \frac{(\zt_1 - \zt_2)^2}{4r_H^2}\right].
    \end{aligned}
\end{equation}
Similar calculations lead to the following expression for the second summation in the second square bracket in \cref{eq:cgc_average_squared_amplitude}
\begin{equation}
    \label{eq:F3_function}
    \begin{aligned}
        & F_3(\zt_1,\zt_2) \equiv  \sum_{i\neq j}^{N_h} \sa{\mu^2(\zt_1-\bt_i)\mu^2(\zt_2-\bt_j)}_{hs,Q_s}\\
        & = \left(\frac{\mu_0^4}{(2\pi)^2 (r_H^2 + R_p^2)}\right)  \left(\frac{N_h(N_h-1)}{r_H^2 + \frac{N_h-2}{N_h}R_p^2}\right) \\
        &\times\exp\left[-\frac{(\zt_1 + \zt_2)^2}{4r_H^2\left(1 + \frac{N_h-2}{N_h}\frac{R_p^2}{r_h^2}\right)} - \frac{(\zt_1 - \zt_2)^2}{4(r_H^2+R_p^2)}\right]. 
    \end{aligned}
\end{equation}
Substituting \cref{eq:F2_function,eq:F3_function} into \cref{eq:cgc_average_squared_amplitude} and performing the Fourier transform, one obtains
\begin{equation}
\label{eq:fourier_transform_squared_amplitude}
\begin{aligned}
    &\int \dd[2]{\bt} \dd[2]{\bt'} e^{-i(\bt-\bt')\cdot\Delta} \sa{\frac{\dd^2\sigma_{\rm dip}}{\dd^2\bt}\frac{\dd^2\sigma_{\rm dip}}{\dd^2\bt'}}_{\mathrm{CGC,hs},Q_s} \\
    & = e^{\sigma_s^2} \{C1\} + \{C2\} + e^{\sigma_s^2} \{DC1\} + \{DC2\},
\end{aligned}
\end{equation}
where
\begin{subequations}
    \begin{equation}
    \begin{aligned}
        \{DC1\} &\equiv \frac{9g^4\mu_0^4 C_F^2}{\pi^2} \frac{e^{-r_H^2 \Delta^2}}{N_h} \\ 
        &\times \left\{ \psi(\rt,\Delta)-\cos\left(\frac{1}{2}\Delta\cdot \rt \right)\psi(0,\Delta)\right\} \\
        &\times \left\{ \psi(\rt',\Delta)-\cos\left(\frac{1}{2}\Delta\cdot \rt' \right)\psi(0,\Delta)\right\},
    \end{aligned}
\end{equation}
\begin{equation}
    \begin{aligned}
        \{DC2\} & \equiv \frac{9g^4\mu_0^4 C_F^2}{\pi^2} \frac{N_h-1}{N_h} e^{-(r_H^2 + R^2)\Delta^2} \\
        & \times \left\{ \psi(\rt,\Delta)-\cos\left(\frac{1}{2}\Delta\cdot \rt \right)\psi(0,\Delta)\right\} \\
        & \times \left\{ \psi(\rt',\Delta)-\cos\left(\frac{1}{2}\Delta\cdot \rt' \right)\psi(0,\Delta)\right\},
    \end{aligned}
\end{equation}
\begin{equation}
    \begin{aligned}
        \{C1\} &\equiv \frac{9g^4\mu_0^4 C_F}{8\pi^4 N_c} \frac{1}{N_h} \int \dd^2\kt \dd^2\kt' e^{-r_H^2(\kt+\kt')^2} \\
        & \times \frac{1}{\left[\left(\kt+\frac{\Delta}{2}\right)^2 + m^2\right]\left[\left(\kt-\frac{\Delta}{2}\right)^2 + m^2\right]}  \\
        & \times \frac{1}{\left[\left(\kt'+\frac{\Delta}{2}\right)^2 + m^2\right]\left[\left(\kt'-\frac{\Delta}{2}\right)^2 + m^2\right]} \\
        &\times \left\{ 2 \cos\left(\frac{1}{2}\Delta\cdot \rt\right) \cos\left(\frac{1}{2}\Delta\cdot \rt'\right) \right. \\
        & \left.- 2 \cos\left(\frac{1}{2}\Delta\cdot \rt\right) e^{i\kt'\cdot \rt'} + e^{i\kt\cdot \rt+i\kt'\cdot \rt'}  \right.\\
        & \left.- 2 \cos\left(\frac{1}{2}\Delta\cdot \rt'\right) e^{i\kt\cdot \rt}  + e^{i\kt\cdot \rt-i\kt'\cdot \rt'}  \right\} ,
    \end{aligned}
\end{equation}
\begin{equation}
    \begin{aligned}
        \{C2\} &\equiv \frac{9g^4\mu_0^4 C_F}{8\pi^4 N_c} \frac{N_h-1}{N_h} \int \dd^2\kt \dd^2\kt' e^{-(r_H^2+R_p^2)(\kt+\kt')^2} \\
        & \times \frac{1}{\left[\left(\kt+\frac{\Delta}{2}\right)^2 + m^2\right]\left[\left(\kt-\frac{\Delta}{2}\right)^2 + m^2\right]}  \\
        & \times \frac{1}{\left[\left(\kt'+\frac{\Delta}{2}\right)^2 + m^2\right]\left[\left(\kt'-\frac{\Delta}{2}\right)^2 + m^2\right]} \\
        &\times \left\{ 2 \cos\left(\frac{1}{2}\Delta\cdot \rt\right) \cos\left(\frac{1}{2}\Delta\cdot \rt'\right) \right. \\
        & \left.- 2 \cos\left(\frac{1}{2}\Delta\cdot \rt\right) e^{i\kt'\cdot \rt'} + e^{i\kt\cdot \rt+i\kt'\cdot \rt'}  \right.\\
        & \left.- 2 \cos\left(\frac{1}{2}\Delta\cdot \rt'\right) e^{i\kt\cdot \rt}  + e^{i\kt\cdot \rt-i\kt'\cdot \rt'}  \right\}. 
    \end{aligned}
\end{equation}

\end{subequations}
Averaging  \cref{eq:fourier_transform_squared_amplitude} over the number of hot spots $N_h$ corresponds to taking the full target average. When this result is convoluted with the wave function overlap, 
the diffractive cross section involving both coherent ($\gamma^*+p\to \jpsim+ p$) and incoherent ($\gamma^*+p\to  \jpsim + Y$) processes reads
\begin{equation}
\label{eq:total_xsection_final}
\begin{aligned}
    \frac{\dd\sigma_{T,L}^{\gamma^*p\to \mathrm{J}/\psi (p,Y)}}{\dd t} &= \frac{\dd\sigma_{T,L}^{\gamma^*p\to \mathrm{J}/\psi p}}{\dd t} + \frac{\dd\sigma_{T,L}^{\gamma^*p\to \mathrm{J}/\psi Y}}{\dd t} \\
    &=  e^{\sigma_s^2}\frac{\dd\sigma_{T,L}^{\rm ci}}{\dd t}|_{DC1} + \frac{\dd\sigma_{T,L}^{\rm ci}}{\dd t}|_{DC2} \\
    &+ e^{\sigma_s^2}\frac{\dd\sigma_{T,L}^{\rm ci}}{\dd t}|_{C1} + \frac{\dd\sigma_{T,L}^{\rm ci}}{\dd t}|_{C2}.
\end{aligned}
\end{equation}
The terms originating from the color-disconnected contribution read
\begin{subequations}
\label{eq:DC_part}
    \begin{equation}
    \begin{aligned}
        \frac{\dd\sigma_{T,L}^{\rm ci}}{\dd t}|_{DC1} = \frac{9C_{T,L}^2}{16\pi} \frac{g^4\mu_0^4C_F^2}{\pi^2} e^{-r_H^2|t|} \mathcal{M}_{T,L}^2 \mathcal{F}(\Lambda), 
    \end{aligned}
    \end{equation}
    \begin{equation}
    \begin{aligned}
        \frac{\dd\sigma_{T,L}^{\rm ci}}{\dd t}|_{DC2} &= \frac{9C_{T,L}^2}{16\pi} \frac{g^4\mu_0^4C_F^2}{\pi^2} e^{-(r_H^2+R_p^2)|t|} \\
        &\times\mathcal{M}_{T,L}^2 \left[1 - \mathcal{F}(\Lambda) \right].
    \end{aligned}
    \end{equation}
\end{subequations}
The remaining terms from the color-connected part are
\begin{subequations}
\label{eq:C_part}
    \begin{equation}
    \begin{aligned}
        \frac{\dd\sigma_{T,L}^{\rm ci}}{\dd t}|_{C1} =  \frac{9C_{T,L}^2}{16\pi} \frac{g^4\mu_0^4C_F}{\pi^2N_c}\mathcal{K}_{T,L}\left(t;r_H^2\right) \mathcal{F}(\Lambda),
    \end{aligned}
    \end{equation}
    \begin{equation}
    \begin{aligned}
        \frac{\dd\sigma_{T,L}^{\rm ci}}{\dd t}|_{C2} = \frac{9C_{T,L}^2}{16\pi} \frac{g^4\mu_0^4C_F}{\pi^2N_c}\mathcal{K}_{T,L}\left(t;r_H^2 + R_p^2\right) \left[1-\mathcal{F}(\Lambda)\right].
    \end{aligned}
    \end{equation}
\end{subequations}
The auxiliary functions $\mathcal{K}_{T,L}$ are defined as
\begin{equation}
\label{ea:K_functions}
\begin{aligned}
    &\mathcal{K}_{T,L}(t;\mathfrak{R}) = \int \dd^2\kt \dd^2\kt' \frac{e^{-\mathfrak{R}(\kt+\kt')^2}}{\left[\left(\kt+\frac{\Delta}{2}\right)^2 + m^2\right]\left[\left(\kt-\frac{\Delta}{2}\right)^2 + m^2\right]} \\
    & \times \frac{\left(\mathcal{X}_{\Delta} - \mathcal{X}_{\kt'}\right)_{T,L}}{\left[\left(\kt'+\frac{\Delta}{2}\right)^2 + m^2\right]\left[\left(\kt'-\frac{\Delta}{2}\right)^2 + m^2\right]}.
\end{aligned}
\end{equation}
where
\begin{subequations}
    \begin{equation}
    \label{eq:U_term}
    \begin{aligned}
        & \mathcal{X}_{\Delta;L} \equiv \frac{1}{2\pi}\int_{|\rt|<r_m} \dd^2\rt \cos\left(\frac{1}{2}\Delta\cdot \rt\right) \\
        \times &  \left\{ AK_0(\varepsilon |\rt|) + \frac{B}{m_c^2}\left[ \frac{9}{2}K_0(\varepsilon |\rt|) + m_c^2|\rt|^2 K_0(\varepsilon |\rt|)  \right.\right. \\
        &\left.\left. - \frac{Q^2|\rt|}{4\varepsilon}K_1(\varepsilon|\rt|) + \frac{1+\cos2\theta_r}{8}(|\rt|^2\Delta^2)K_0(\varepsilon|\rt|)\right]\right\},
    \end{aligned}
    \end{equation}
    \begin{equation}
    \begin{aligned}
       \mathcal{X}_{\Delta;T} &\equiv \int_{|\rt|<r_m} \frac{\dd^2\rt}{2\pi} \cos\left(\frac{1}{2}\Delta\cdot \rt\right) 
         \left\{ AK_0(\varepsilon |\rt|) \right. \\
        &\left. + \frac{B}{m_c^2}\left[ \frac{7}{2}K_0(\varepsilon |\rt|) + m_c^2|\rt|^2 K_0(\varepsilon |\rt|)  \right.\right. \\
        &\left.\left. - \left(\frac{Q^2}{4\varepsilon} + \varepsilon\right)|\rt|K_1(\varepsilon|\rt|) \right.\right. \\
        &\left.\left.+ \frac{1+\cos2\theta_r}{8}(|\rt|^2\Delta^2)K_0(\varepsilon|\rt|)\right]\right\},
    \end{aligned}
    \end{equation}
    
\end{subequations}
\begin{subequations}
    \begin{equation}
    \label{eq:V_term}
    \begin{aligned}
        & \mathcal{X}_{\kt;L} \equiv \frac{1}{2\pi}\int_{|\rt|<r_m} \dd^2\rt e^{i\kt\rt} \\
        \times &  \left\{ AK_0(\varepsilon |\rt|) + \frac{B}{m_c^2}\left[ \frac{9}{2}K_0(\varepsilon |\rt|) + m_c^2|\rt|^2 K_0(\varepsilon |\rt|)  \right.\right. \\
        &\left.\left. - \frac{Q^2|\rt|}{4\varepsilon}K_1(\varepsilon|\rt|) + \frac{1+\cos2\theta_r}{8}(|\rt|^2\Delta^2)K_0(\varepsilon|\rt|)\right]\right\},
    \end{aligned}
    \end{equation}
    and
    \begin{equation}
    \begin{aligned}
        & \mathcal{X}_{\kt;T} \equiv \frac{1}{2\pi}\int_{|\rt|<r_m} \dd^2\rt e^{i\kt\rt} \\
        \times &  \left\{ AK_0(\varepsilon |\rt|) + \frac{B}{m_c^2}\left[ \frac{7}{2}K_0(\varepsilon |\rt|) + m_c^2|\rt|^2 K_0(\varepsilon |\rt|)  \right.\right. \\
        &\left.\left. - \left(\frac{Q^2}{4\varepsilon} + \varepsilon\right)|\rt|K_1(\varepsilon|\rt|) + \frac{1+\cos2\theta_r}{8}(|\rt|^2\Delta^2)K_0(\varepsilon|\rt|)\right]\right\}.
    \end{aligned}
    \end{equation}
\end{subequations}

Different from the corresponding result in Ref.~\cite{Demirci:2022wuy}, the total exclusive  (coherent+incoherent) $\mathrm{J}/\psi$ production cross section (\ref{eq:total_xsection_final}) involves the dependence on both the $Q_s$ and $N_h$ fluctuations, as well as the dependence on the relativistic correction (terms proportional to $B/m_c^2$). From \cref{eq:total_xsection_final}, we see that the $Q_s$ fluctuations affect only the single-hot-spot contributions ($DC1$ and $C1$).

To estimate the effect of the $Q_s$ fluctuations on the incoherent cross section, let us ignore the $N_h$ fluctuations and subtract the coherent contribution from \cref{eq:total_xsection_final} as was done in Ref.~\cite{Demirci:2022wuy}. We denote by $R_{X}, X \in \{DC1,DC2,C1,C2\}$, the ratios of the 4 contributions to the incoherent cross section in the case where   the $Q_s$ fluctuations are not included. As such, $\sum_X R_X=1$. From Ref.~\cite{Demirci:2022wuy} we get that $0<R_{C1}\sim R_{C2} <1$, $R_{DC2}<0$ and $R_{DC}>0$ (and this $DC1$ component can be larger than unity).
Note that the $R_{DC2}$ component can be negative as we have subtracted the coherent cross section.
As such, the $Q_s$ fluctuations increase the incoherent cross section by a $t$-dependent factor
%
\begin{equation}
\label{eq:ratio_sigma}
\begin{aligned}
    &\Phi_{\sigma_s}  \equiv  \frac{(\dd\sigma/\dd t)^{\rm incoh}_{\sigma_s>0}}{(\dd\sigma/\dd t)^{\rm incoh}_{\sigma_s=0}}\\ 
    & = 1 +(e^{\sigma_s^2}-1)\left(R_{DC1} + R_{C1} + \frac{1}{N_h}\underbrace{\frac{(\dd\sigma/\dd t)^{\rm  \gamma^*p\to \mathrm{J}/\psi p }}{(\dd\sigma/\dd t)^{\rm incoh}_{\sigma_s=0}}}_{ \mathcal{R}}\right).
\end{aligned}
\end{equation}
Here $\mathcal{R}$ is the incoherent-to-coherent cross section ratio in the case where there are no $Q_s$ fluctuations\footnote{Recall that the $Q_s$ fluctuations have no effect on the coherent production.}.
Note that the dependence on coherent cross section appears in \cref{eq:ratio_sigma} as we are calculating the modification factor for incoherent cross section, which is calculated by subtracting the coherent contribution from the total diffractive cross section \cref{eq:total_xsection_final}.

Let us next consider the effect of $Q_s$ fluctuations in three different kinematical regimes. 
\begin{enumerate}
\item At large $|t| \gg 1/r_H^2 $,  color charge fluctuations, as described by the the $C$ terms, dominate. Furthermore, the coherent cross section is subleading in this regime, so we can ignore the coherent-over-incoherent ratio $\mathcal{R}$. In this case, $R_{DC1} + R_{C1} \approx R_{C1}$ , $1 = \sum R_X \approx R_{C1} + R_{C2}$ and $R_{C1} \sim R_{C2}$. Therefore, $ 1< \Phi_{\sigma_s} \approx 1 + (e^{\sigma_s^2}-1) R_{C1} < e^{\sigma_s^2}$. In other words, in this regime, the incoherent cross section is enhanced by a factor less than $e^{\sigma_s^2}$.
\item At intermediate $|t|\sim 1/r_H^2$ where the hot spot fluctuations ($DC$ terms) dominate and the coherent contribution is still significantly smaller than the incoherent one, $R_{DC1}\simeq 1$ and $\Phi_{\sigma_s} \approx 1 + (e^{\sigma_s^2}-1) R_{DC1} \simeq e^{\sigma_s^2}$.  This means that the $Q_s$ fluctuations increase the incoherent cross section approximatively by a factor of  $e^{\sigma_s^2}$. 
\item Finally, at small $|t| \ll 1/r_H^2$, the color charge fluctuations again dominate over  hot spot fluctuations. However, in this range, the coherent contribution becomes important, and the coherent-over-incoherent ratio $\mathcal{R}\sim 10$~\cite{H1:2013okq}. In this case, the enhancement  $\Phi_s$ depends strongly on $\mathcal{R}$ and $N_h$. For $\mathcal{R}/N_h > 1$, the enhancement can be significant, with $\Phi_{\sigma_s} > e^{\sigma_s^2}$. 
\end{enumerate}
To summarize, the $Q_s$ fluctuations enhance the incoherent cross section by a $|t|$-dependent factor, which is typically of the order of $e^{\sigma_s^2}$.    

The dependence on the $N_h$ fluctuations (or on $\Lambda$ that controls the mean number of hot spots in \eqref{eq:hotspot_distribution}) also factorizes in both the single-hot-spot (as $\mathcal{F}(\Lambda)$) and the double-hot-spot (as $1-\mathcal{F}(\Lambda)$) terms. At large  $|t|\gtrsim 1/r_H^2$,  when the coherent contribution can be negligible, the ratio of the incoherent cross section involving the $N_h$ fluctuations versus the one where the number of hot spots is fixed to $N_h^f $ reads 
\begin{multline}
    \Phi_{\Lambda} \equiv N_h^{f}\mathcal{F}(\Lambda) (R_{DC1} + R_{C1}) \\ + \frac{N_h^{f}}{N_h^{f}-1}[1-\mathcal{F}(\Lambda)](R_{DC2} + R_{C2}) \\
    = \frac{N_h^{f}[1-\mathcal{F}(\Lambda)]}{N_h^{f}-1} \left[1 + \frac{N_h^f\mathcal{F}(\Lambda) -1  }{1-\mathcal{F}(\Lambda)}(R_{DC1} + R_{C1}) \right].
\end{multline}
In particular, at $|t|\sim 1/r_H^2$ we find $\Phi_{\Lambda} \simeq N_h^f\mathcal{F}(\Lambda)$, which implies that the modification to the incoherent cross section depends strongly on the value of $N_h^f\mathcal{F}(\Lambda)$. 

To estimate the phenomenological impact of the $N_h$ fluctuations, we fix the number of hot spots in the reference to a typical value $N_h^f=3$. We obtain $\Phi_{\Lambda}= 1.01\dots1.37$, where the band is obtained by varying $\overline{N_h}$ in the range $\overline{N_h}=3\dots 4$. For this specific choice, the $N_h$ fluctuations have a modest effect, as $\Phi_\Lambda$ is close to unity.

\section{Bayesian Parameter Estimation}
\label{sec:bayesian}

The hotspot model contains several non-perturbative parameters that must be extracted from data. In this analysis, their values are constrained using Bayesian inference. In particular, the unknown parameters $\mathbf{\Theta}$ are estimated from their posterior distribution, which is determined using Bayes's theorem as
\begin{equation}
\label{eq:bayes_theorem}
    \mathbb{P}\left(\mathbf{\Theta|Y_{\rm exp}}\right) \propto \mathbb{P}\left(\mathbf{Y_{\rm exp}|\Theta}\right) \mathbb{P}(\mathbf{\Theta}).
\end{equation}
The distribution $\mathbb{P}(\mathbf{\Theta})$ in the above equation encodes the prior knowledge on the model parameters. Here we choose the flat multivariate distribution with bounds corresponding to physically motivated parameter ranges. 

The function $\mathbb{P}\left(\mathbf{Y_{\rm exp}|\Theta}\right)$ is the likelihood of the model calculations with parameters $\mathbf{\Theta}$ to match the experimental data. It is chosen to be the multivariate Gaussian distribution, for which the log-likelihood reads
\begin{equation}
    \label{eq:log-likelihood}
    \begin{aligned}
        &\log \left[ \mathbb{P}\left(\mathbf{Y_{\rm exp}|\Theta}\right) \right] = -\frac{1}{2}\log\left[(2\pi)^n {\rm det}\Sigma\right]\\
        &- \frac{1}{2} \left[\mathbf{Y(\Theta) - Y_{\rm exp}}\right]^{T}\Sigma^{-1}\left[\mathbf{Y(\Theta) - Y_{\rm exp}}\right],    
    \end{aligned}
\end{equation}
where $\mathbf{Y(\Theta)}$ is the model result calculated using parameters $\mathbf{\Theta}$, $\mathbf{Y_{\rm exp}}$ represents the experimental data, and $n = {\rm dim}(\mathbf{Y_{\rm exp}})$ is the number of experimental data points used in the analysis. The matrix $\Sigma = \Sigma_{\rm exp} + \Sigma_{\rm model}$ is the $(n\times n)$ covariance matrix encoding the uncertainties from both experimental and model analyses. In the current study, the experimental part is assumed to be diagonal: 
\begin{equation}
    \label{eq:covariance_matrix_exp}
    \Sigma_{\rm exp} = \rm diag(\sigma_1^2,\sigma_2^2\hdots,\sigma_n^2),
\end{equation}
where $\sigma_i,~(k=1,\hdots, n)$ are experimental errors. We use the H1 2006--2007 dataset~\cite{H1:2013okq} for exclusive $\mathrm{J}/\psi$ photoproduction differential cross sections $\dd\sigma/\dd t$ in both coherent and incoherent channels at the kinematics $Q^2 = 0.1~\gev^2$ and $W = 78~\gev$ for the current analysis.  

The model uncertainties are assessed using the covariance matrix obtained from trained Gaussian Process Emulators (GPEs)~\cite{williams2006gaussian}, which act as computationally efficient surrogates for the full theoretical calculations. These emulators enable rapid evaluations of model predictions across the parameter space, significantly reducing computational cost. The parameter space is explored using the Markov Chain Monte Carlo (MCMC) method, implemented with the Python-based \texttt{emcee} package~\cite{2010CAMCS...5...65G,2013PASP..125..306F} with multiple random walkers. 
Training of the GPEs is performed using a set of points selected via the maximin Latin hypercube sampling strategy, which ensures a well-distributed and representative coverage of the parameter space.%

The set of free parameters $\bf \Theta$ to be constrained by the H1 data on the exclusive photoproduction of the \jpsi involves:

\begin{itemize}
    \item $g\mu_0$: the average amount of color charge in the hot spot, 
    \item $m^2$: the IR regulator in \cref{eq:Green_function},
    \item $R_p$: parameter controlling the proton size in \cref{eq:hotspot_distribution},
    \item $r_H$: parameter controlling the hot spot size in \cref{eq:hotspot_profile},
    \item $\overline{N}_h$: the average number of hot spots when  fluctuations in this quantity are taken into account. In the absence of such fluctuation, the corresponding free parameter reduces to the number of hot spots $N_h$. In the previous analyses~\cite{Demirci:2021kya,Demirci:2022wuy}, $N_h$ is fixed at the value $N_h=3$. In the present work, however, we treat it as a free parameter. Specifically, we sample $N_h$ within the hypercube and define $\lfloor N_h + 0.5\rfloor$ as the instantaneous number of hot spots.
    \item $\sigma_s$: the magnitude of fluctuations in the saturation scale $Q_s$ as parametrized in \cref{eq:Qs_fluct}.  When such fluctuations are neglected, the saturation scale is assumed to be fixed at its average value across all events, corresponding to $\sigma_s=0$.
\end{itemize}
The charm quark mass is set to $m_c=1.4~\gev$ as in Ref.~\cite{Lappi:2020ufv},
consistently with the extraction of the applied NRQCD long-distance matrix elements~\cite{Bodwin:2007fz}.

\section{Numerical results}
\label{sec:numerics}

\begin{figure*}[tb]
  \begin{minipage}[b]{.52\linewidth}
    \centering
    \includegraphics[width=\linewidth]{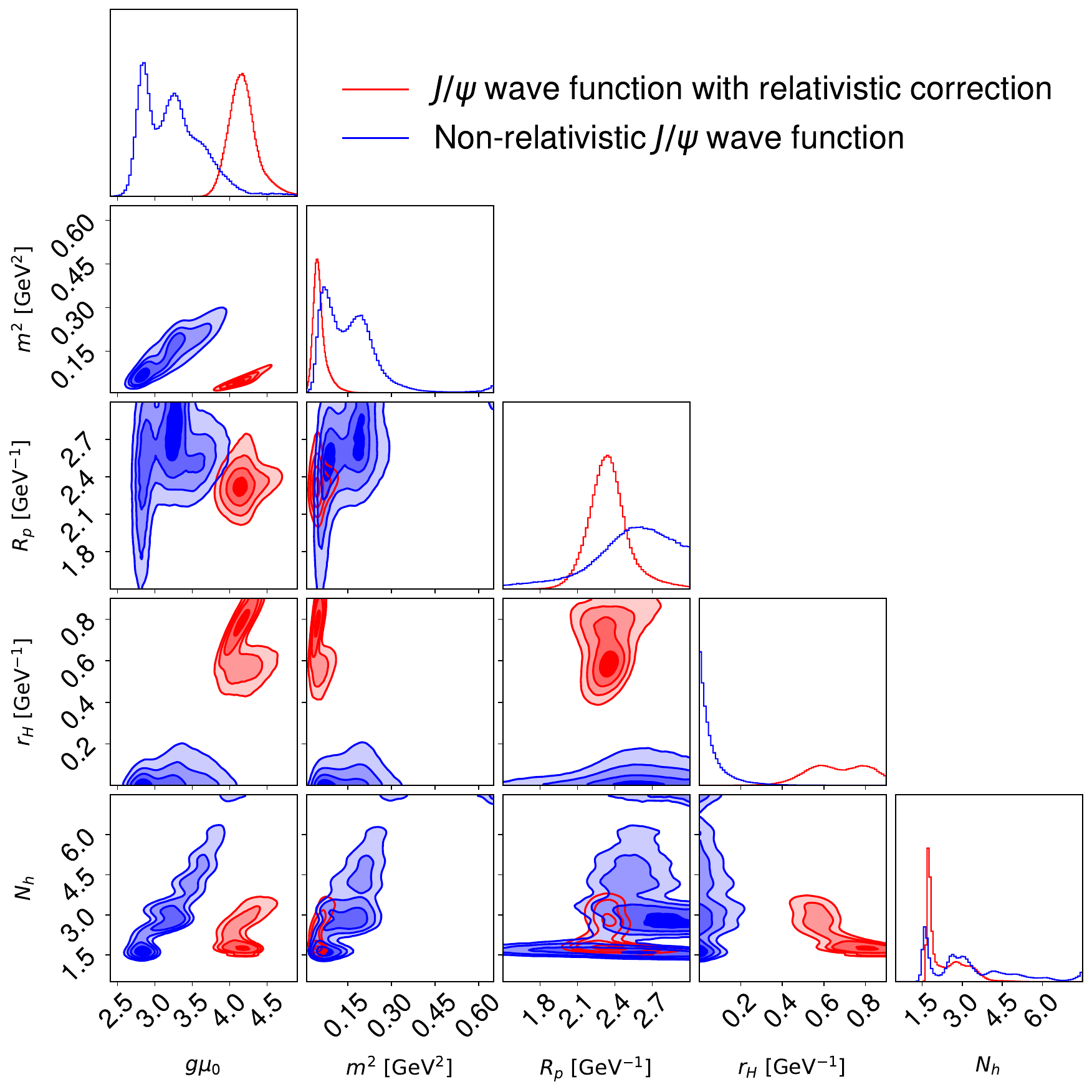}
    \captionof{figure}{Posterior distribution of parameters in the scheme without $Q_s$ or $N_h$ fluctuations.
    }
    \label{fig:corner_plot_wo_Nh_wo_Qs}
  \end{minipage}\hspace{0.1em}
  \begin{minipage}[b]{.42\linewidth}
    \centering
    {\renewcommand{\arraystretch}{2}
    \setlength{\tabcolsep}{0.2em}
    \begin{tabular}{ *{1}{c}*{1}{c}*{1}{c}*{1}{c} }
    \hline
      {Parameters} & {Prior range} & {MAP~(R)} & {MAP~(NR)}  \\
      \hline
      $g\mu_0$ & $[2.4,4.9]$  & $\cpm{4.037}{0.225}{0.559}$ & $\cpm{2.857}{0.198}{1.130}$\\
      $m^2~[\rm GeV^2]$ & $[0.008,0.65]$ & $\cpm{0.037}{0.023}{0.067}$& $\cpm{0.076}{0.048}{0.242}$\\
      $R_p~[\rm GeV^{-1}]$ & $[1.5,3.0]$ & $\cpm{2.366}{0.356}{0.344}$ & $\cpm{2.310}{0.491}{0.69}$\\
      $r_H~[\rm GeV^{-1}]$ & $[0,0.9]$ & $\cpm{0.699}{0.261}{0.201}$ & $\cpm{0.001}{0.001}{0.219}$\\
      ${N_h}$ & $[0.5, 7.5]$ & $\cpm{1.760}{0.138}{1.907}$ & $\cpm{1.608}{0.213}{4.797}$\\
      \hline
      \multicolumn{2}{c}{$\chi^2/\rm{d.o.f}$} & $1.63$ & $4.14$\\
      \hline
    \end{tabular}
    \vspace{2.7em}
    \captionof{table}{Parameter estimation from the posterior distribution shown in \cref{fig:corner_plot_wo_Nh_wo_Qs}.} 
    \label{tab:params_wo_Nh_wo_Qs}
    }
  \end{minipage}
\end{figure*}

\begin{figure*}[tb]
    \centering
    \includegraphics[width=\textwidth]{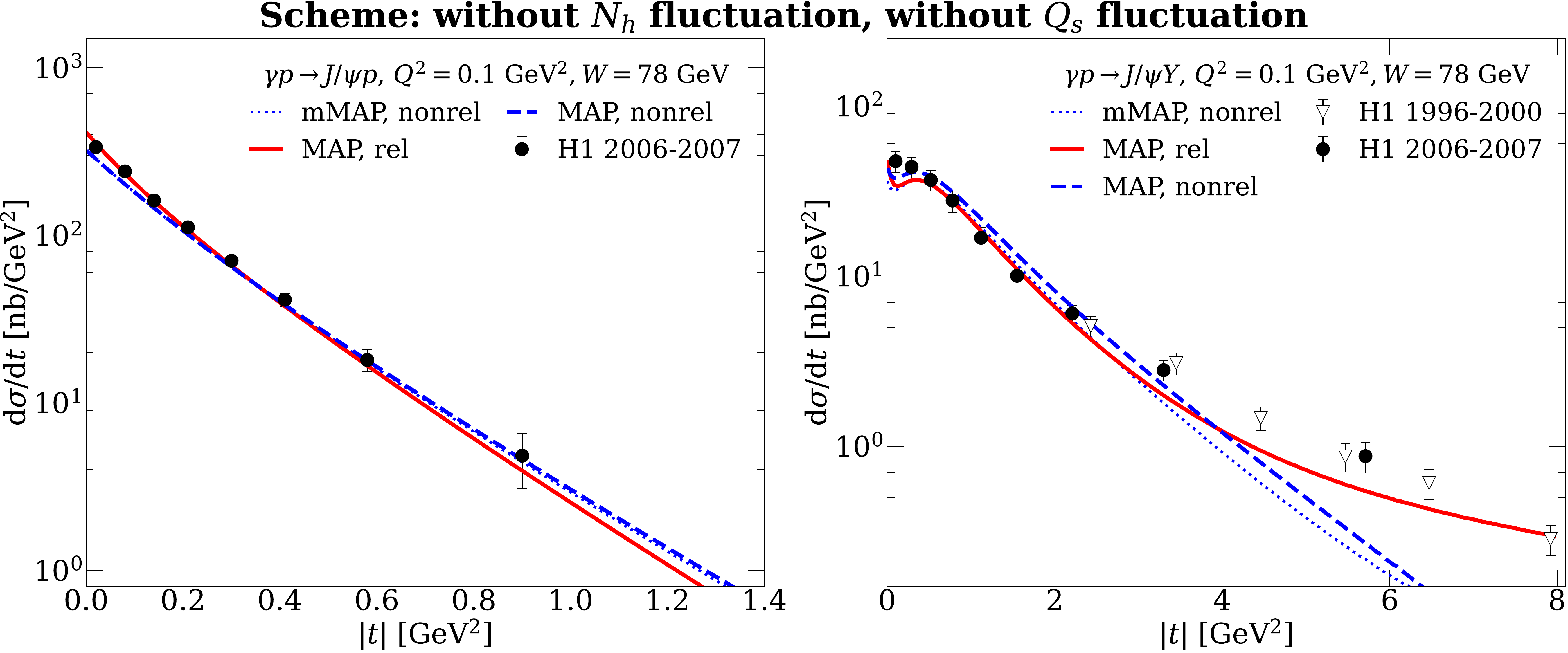}
    \caption{
    Fit results  compared to the H1 data~\cite{H1:2013okq,H1:2003ksk} up to $|t|=8~\gev^2$.  For the fully non-relativistic setup, the $\rm mMAP$ parameter set  is that of $\rm MAP$ but with $r_H = 0.2 ~\gev^{-1}$. 
    }
     \label{fig:datacomparison_wo_Nh_wo_Qs}
\end{figure*}

The first Bayesian fit is performed under the assumption that both $N_h$ and $Q_s$ fluctuations are neglected, i.e. in the non-relativistic limit it corresponds to the setup studied in Ref.~\cite{Demirci:2022wuy}.
Posterior distributions obtained in the non-relativistic limit, and when the first relativistic correction is included, are shown in \cref{fig:corner_plot_wo_Nh_wo_Qs} and the corresponding maximum a posteriori (MAP) parameters in \cref{tab:params_wo_Nh_wo_Qs}. Comparison to the H1 data~\cite{H1:2003ksk,H1:2013okq} is shown in \cref{fig:datacomparison_wo_Nh_wo_Qs}.

In the fully non-relativistic (NR or nonrel) limit, the resulting cross sections computed using the maximum a posteriori (MAP) values yield poor agreement with the incoherent cross section   data at large $|t| \gtrsim 4$~$\gev^2$, as indicated by large reduced chi-squared values ($\chi^2/{\rm dof} = 4.14$). On the other hand, the description of the available coherent and incoherent cross section data at smaller $|t|$ is good. 
Note that the corresponding MAP hot spot size is unphysically small ($r_H\approx0 ~\gev^{-1}$), implying that the color charge density collapses into a sharply peaked, delta-function-like structure. However, the field of an individual point charge should extend to a distance of order $1/m$~\cite{Demirci:2022wuy}. 
When this parameter is adjusted to a more physically reasonable value, $r_H = 0.2~\gev^{-1}$, the resulting predictions (denoted as mMAP in \cref{fig:datacomparison_wo_Nh_wo_Qs}) still yield a satisfactory description of the data except at high-$|t|$. As the color field obtained from the unphysically small color charge density already extends to distances $\sim 1/m$, adjusting the hot spot size here has only a small effect at the cross-section level.

\begin{figure*}[tb]
    \centering
    \includegraphics[width=\linewidth]{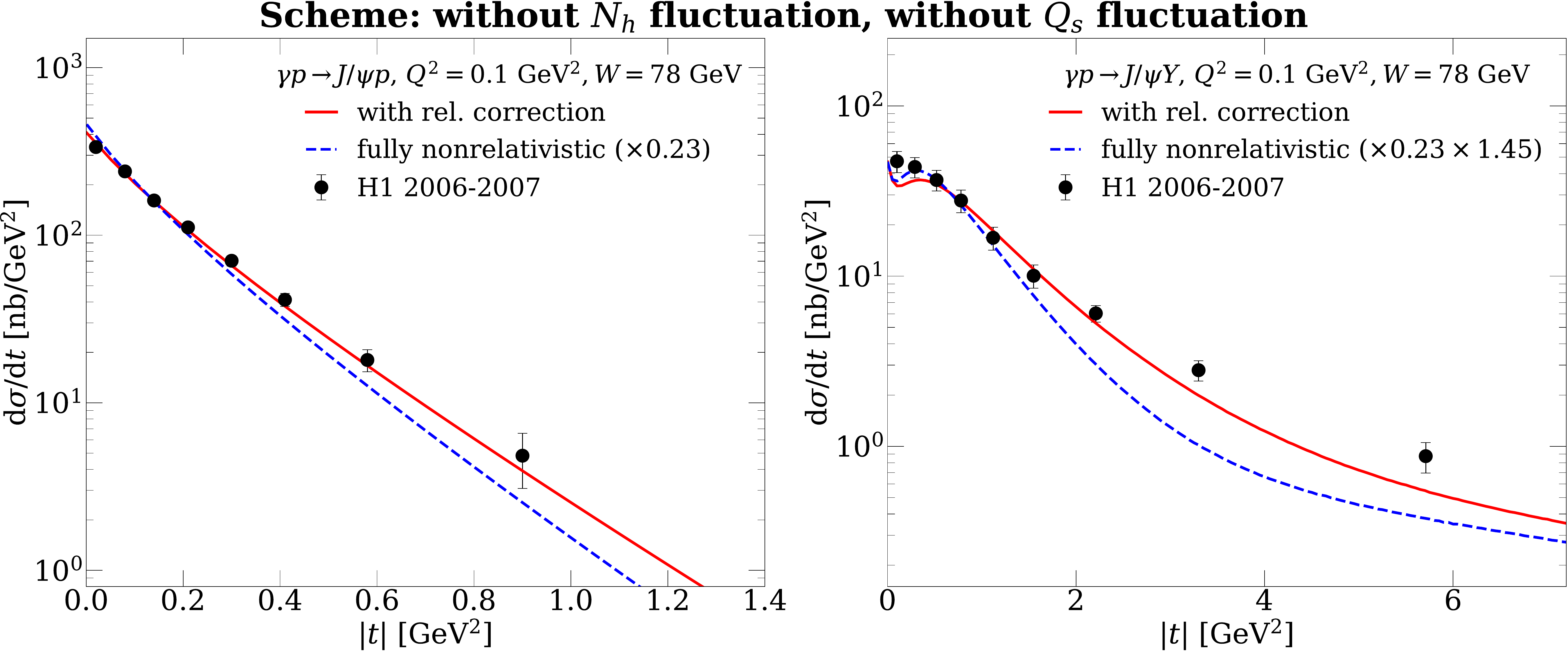}
    \caption{
    Coherent and incoherent cross sections calculated in the non-relativistic limit and with the first relativistic correction included. 
    The scaling factor $0.23$ is to scale the corresponding coherent curve to the experimental data. 
}   
    \label{fig:rel_effect}
\end{figure*}

This finding stands in contrast to the normalization mismatch argument given in Ref.~\cite{Demirci:2022wuy}, where a multiplicative factor of approximately $2.5$ was proposed to reconcile the incoherent cross section with experimental data, attributed to the absence of additional sources of fluctuations. However, a direct comparison between the MAP (or mMAP) parameter set and the manually selected parameters used in Ref.~\cite{Demirci:2022wuy} suggests that the claimed mismatch is likely an artifact of parameter selection, rather than a fundamental issue with missing fluctuations. As an example, in this fit, the number of hot spots peaks at $N_h=2$, while it was chosen to be $N_h=3$ in Refs.~\cite{Demirci:2021kya,Demirci:2022wuy}. 

When relativistic corrections are included, a good description of the H1 data with $\chi^2/{\rm dof}=1.63$ is obtained at all $t$. Furthermore, parameter values are physically better motivated; in particular, the hot spot size can now be constrained to a finite value. 
The main effect of the relativistic correction is to significantly suppress the cross section, as reflected by the larger color charge density needed to compensate for this correction. This effect is expected based on Ref.~\cite{Lappi:2020ufv}, and can be seen from the wave function overlap \eqref{eq:wf_overlap} noticing that the non-relativistic contribution and the relativistic correction come with opposite signs ($A>0,B<0$). 

The relativistic correction is also found to have a significant effect on the shape of the incoherent $t$ spectrum at large $|t|$, as shown in \cref{fig:datacomparison_wo_Nh_wo_Qs} (right panel).
This correction  changes the spectrum from its exponential shape to a power-law behavior at  $|t|\gtrsim 3.5 ~\gev^2$. However, this effect actually comes from the magnitude of $r_H$. As discussed in Ref.~\cite{Demirci:2022wuy}, the hot spot fluctuations are probed at $|t| \sim 1/r_H^2$, and beyond this scale, there is a transition to the power-law (more precisely $\ln|t|/t^2$, see Ref.~\cite{Demirci:2022wuy}) behavior originating from the  color charge fluctuations. Therefore, there should be a transition to the power-law behavior in the non-relativistic setup also, but this transition takes place at much larger $|t|$.

Before quantifying the impact of the $Q_s$ and number of hot spot fluctuations, let us analyze in more detail the effect of the relativistic correction to the \jpsi wave function. As already seen above and in Ref.~\cite{Lappi:2020ufv}, the relativistic correction significantly suppresses the photoproduction cross section. In order to quantify the effect of the relativistic correction on coherent and incoherent cross sections separately, we calculate these cross sections using the MAP parameters corresponding to the non-relativistic setup listed in \cref{tab:params_wo_Nh_wo_Qs}, with and without the relativistic correction included in the model. Then, we separately scale the coherent and incoherent cross sections obtained in the non-relativistic limit such that they approximately match the calculation with the relativistic correction included. The obtained  spectra and the determined scaling factors are shown in Fig.~\ref{fig:rel_effect}  

We find that the scaling factors required for the coherent and incoherent cross sections differ significantly. In particular, the first relativistic correction suppresses the coherent cross section much more than the incoherent one.
This is because the relativistic correction puts more weight on smaller dipoles that can more effectively resolve substructure fluctuations enhancing the incoherent cross section. 
As such, the relativistic correction can be accounted for as a missing normalization factor of about $2.5$ for the incoherent cross section, as was discussed in Ref.~\cite{Demirci:2022wuy}. 
Furthermore, as discussed above, including this relativistic correction is essential for describing the exponential-to-power-law transition observed in the experimental data at $t\gtrsim 3.5~\gev^2$.    

\begin{figure*}[ht!]
  \begin{minipage}[b]{.55\textwidth}
    \centering
    \includegraphics[width=\linewidth]{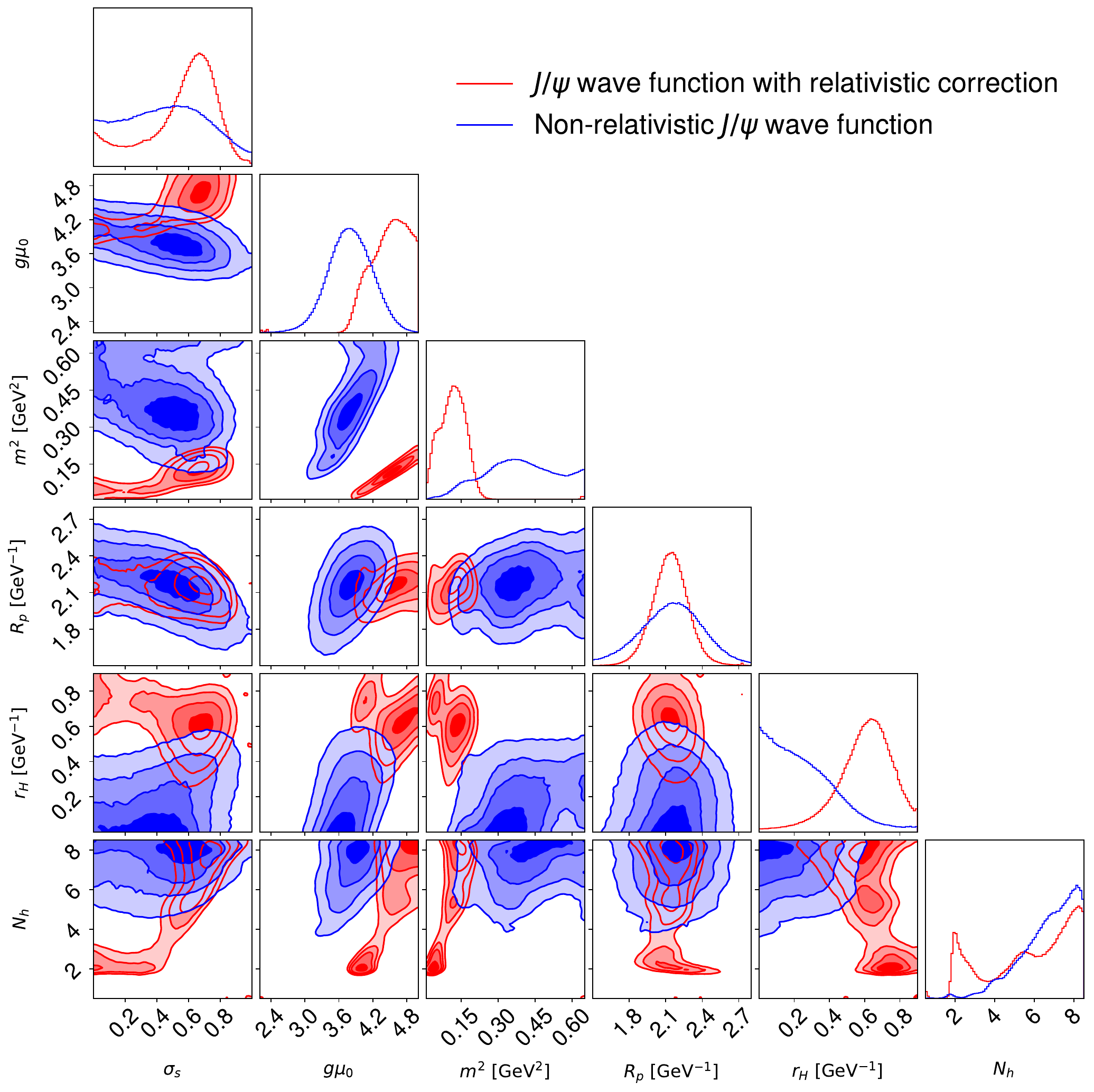}
    \captionof{figure}{Posterior distribution of parameters in the scheme with $Q_s$ fluctuations and without $N_h$ fluctuations (variable $N_h$).}
    \label{fig:corner_plot_wo_Nh_w_Qs}
  \end{minipage}\hspace{0.1em}
 \begin{minipage}[b]{.42\linewidth}
    \centering
    {\renewcommand{\arraystretch}{2}
    \setlength{\tabcolsep}{0.2em}
    \begin{tabular}{ *{1}{c}*{1}{c}*{1}{c}*{1}{c} }
    \hline
      {Parameters} & {Prior range} & {MAP~(R)} & {MAP~(NR)}  \\
      \hline
      $\sigma_s$ & $[0.0, 1.0]$  & $\cpm{0.595}{0.583}{0.252}$ & $\cpm{0.586}{0.586}{0.290}$\\
      $g\mu_0$ & $[2.2, 5.0]$  & $\cpm{4.517}{0.577}{0.483}$ & $\cpm{3.558}{0.501}{1.014}$\\
      $m^2~[\rm GeV^2]$ & $[0.008,0.65]$ & $\cpm{0.107}{0.088}{0.095}$& $\cpm{0.191}{0.046}{0.459}$\\
      $R_p~[\rm GeV^{-1}]$ & $[1.5,2.8]$ & $\cpm{2.108}{0.264}{0.323}$ & $\cpm{2.109}{0.479}{0.521}$\\
      $r_H~[\rm GeV^{-1}]$ & $[0.0, 0.9]$ & $\cpm{0.627}{0.328}{0.273}$ & $\cpm{0.008}{0.008}{0.613}$\\
      ${N_h}$ & $[0.5, 8.5]$ & $\cpm{5.098}{3.076}{3.402}$ & $\cpm{8.186}{4.132}{0.314}$\\
      \hline
      \multicolumn{2}{c}{$\chi^2/\rm{d.o.f}$} & $1.24$ & $1.73$\\
      \hline
    \end{tabular}
    \vspace{2.7em}
    \captionof{table}{Parameter estimation from the posterior distribution in \cref{fig:corner_plot_wo_Nh_w_Qs}.  }
    \label{tab:params_wo_Nh_w_Qs}
    }
  \end{minipage}
\end{figure*}

\begin{figure*}[ht!]
    \centering
    \includegraphics[width=\textwidth]{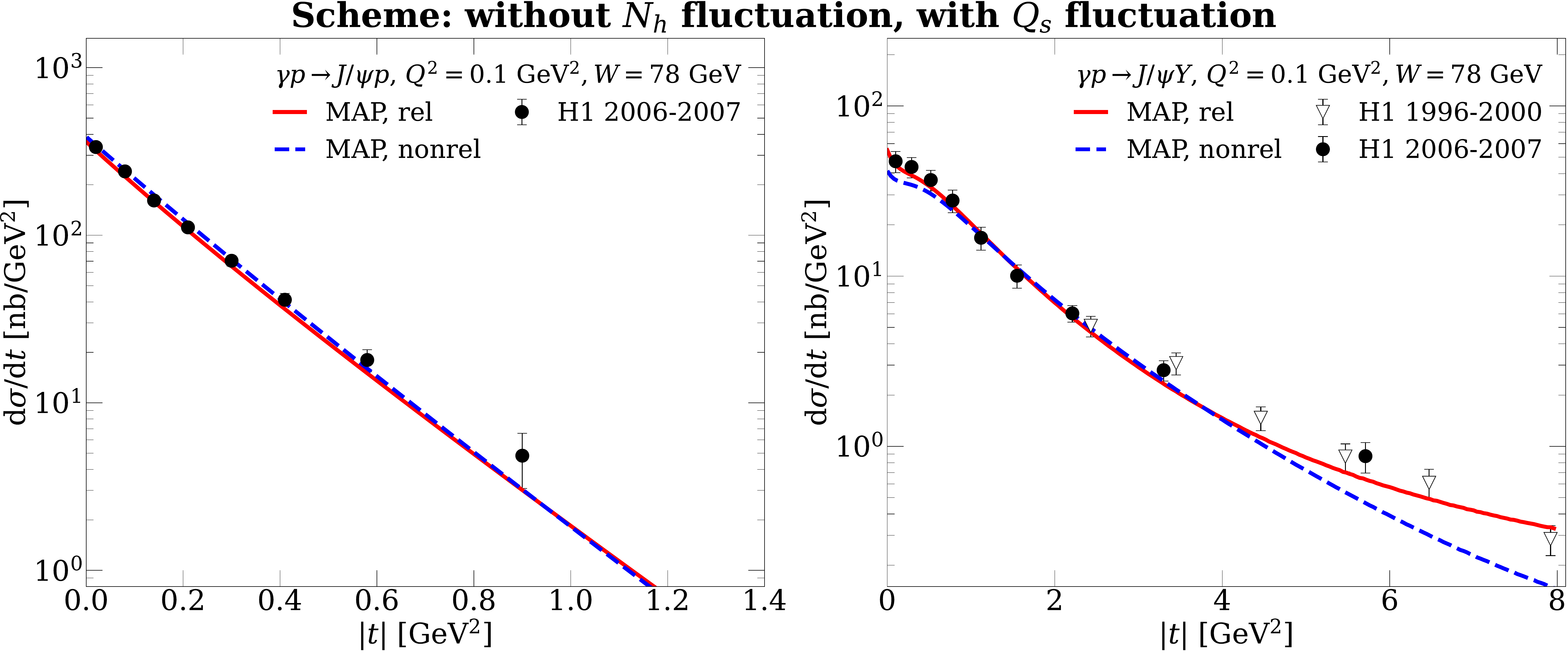}
    \caption{
    Fit result compared to the H1 data~\cite{H1:2013okq,H1:2003ksk}.
    }
     \label{fig:datacomparison_wo_Nh_w_Qs}
\end{figure*}

Next we move beyond the setup used in Ref.~\cite{Demirci:2022wuy} and include the $Q_s$ fluctuations into the model. 
The posterior distribution obtained for the model parameters in this case is shown in
\cref{fig:corner_plot_wo_Nh_w_Qs}, with the MAP parameters summarized in \cref{tab:params_wo_Nh_w_Qs}. Comparison to the HERA data is shown in \cref{fig:datacomparison_wo_Nh_w_Qs}. The resulting MAP parameter sets yield improved fits to the H1 data across both modeling schemes -- with and without relativistic corrections -- as evidenced by the consistently lower reduced $\chi^2$ values.  Notably, for the incoherent cross-section, the MAP fit incorporating relativistic corrections provides a more accurate description of the data, particularly in the small- and large-$|t|$ regions. From the MAP fit with the relativistic correction, we can conclude that by including both the relativistic correction and the event-by-event $Q_s$ fluctuations allows a good description of the cross-section data in the whole $|t|$ range, provided that $|t|$ is not very large.

Similarly to the first case, the MAP value for the hot spot size $r_H$ is peaked at zero in the non-relativistic limit. Consequently, the transition to the power-law behavior is again not seen at $|t|\sim 3.5~ \gev^2$ in this setup. However, the corresponding marginal distribution of $r_H$ is considerably flatter than the one from the setup without $Q_s$ fluctuations.  

We also consider the number of hot spots $N_h$ to be a free parameter. However, we find that it is not well constrained by the H1 data. Similar results have been obtained in Ref.~\cite{Mantysaari:2022ffw}. This is because the number of hot spots is correlated with the magnitude of $Q_s$ fluctuations, a correlation which is especially clear in the setup with relativistic corrections. With large $Q_s$ fluctuations, some hot spots will always have a significant reduction in their density, and as such, the effective number of hot spots is reduced.

\begin{figure*}[tb]
  \begin{minipage}[b]{.55\linewidth}
    \centering
    \includegraphics[width=\linewidth]{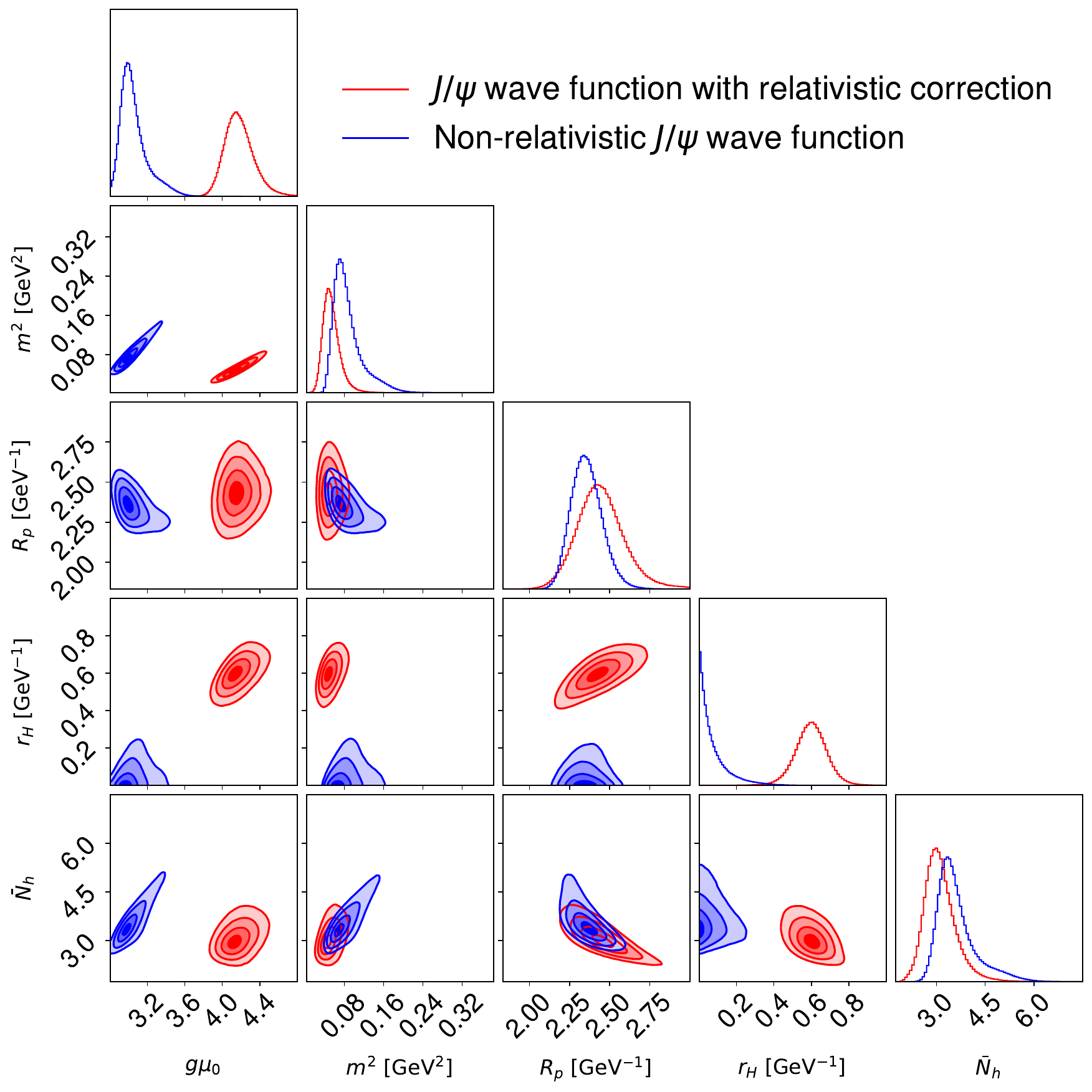}
   
    \captionof{figure}{Posterior distribution of parameters in the scheme without $Q_s$ fluctuation and with $N_h$ fluctuation.}%
    \label{fig:corner_plot_w_Nh_wo_Qs}
  \end{minipage}\hspace{0.1em}
  \begin{minipage}[b]{.42\linewidth}
    \centering
    {\renewcommand{\arraystretch}{2}
    \setlength{\tabcolsep}{0.2em}
    \begin{tabular}{ *{1}{c}*{1}{c}*{1}{c}*{1}{c} }
    \hline
      {Parameters} & {Prior range} & {MAP~(R)} & {MAP~(NR)}  \\
      \midrule      \hline
      $\sigma_s$ & No $Q_s$ fluctuation  & $0$ & $0$\\
      $g\mu_0$ & $[2.8, 4.8]$  & $\cpm{4.116}{0.226}{0.38}$ & $\cpm{2.946}{0.125}{0.439}$\\
      $m^2~[\rm GeV^2]$ & $[0.001,0.86]$ & $\cpm{0.045}{0.022}{0.044}$& $\cpm{0.063}{0.021}{0.090}$\\
      $R_p~[\rm GeV^{-1}]$ & $[1.6,3.0]$ & $\cpm{2.458}{0.302}{0.289}$ & $\cpm{2.382}{0.205}{0.167}$\\
      $r_H~[\rm GeV^{-1}]$ & $[0.0, 1.0]$ & $\cpm{0.595}{0.17}{0.173}$ & $\cpm{0.002}{0.001}{0.246}$\\
      $\overline{N}_h$ & $[1.2, 7.5]$ & $\cpm{2.925}{0.691}{1.207}$ & $\cpm{3.303}{0.623}{1.696}$\\
      \hline
      \multicolumn{2}{c}{$\chi^2/\rm{d.o.f}$} & $1.04$ & $1.94$\\
      \hline
    \end{tabular}
    \vspace{2.7em}
    \captionof{table}{Parameter estimation from the posterior distribution in \cref{fig:corner_plot_w_Nh_wo_Qs}.  }
    \label{tab:params_w_Nh_wo_Qs}
    }
  \end{minipage}
\end{figure*}

\begin{figure*}[ht!]
    \centering
    \includegraphics[width=\textwidth]{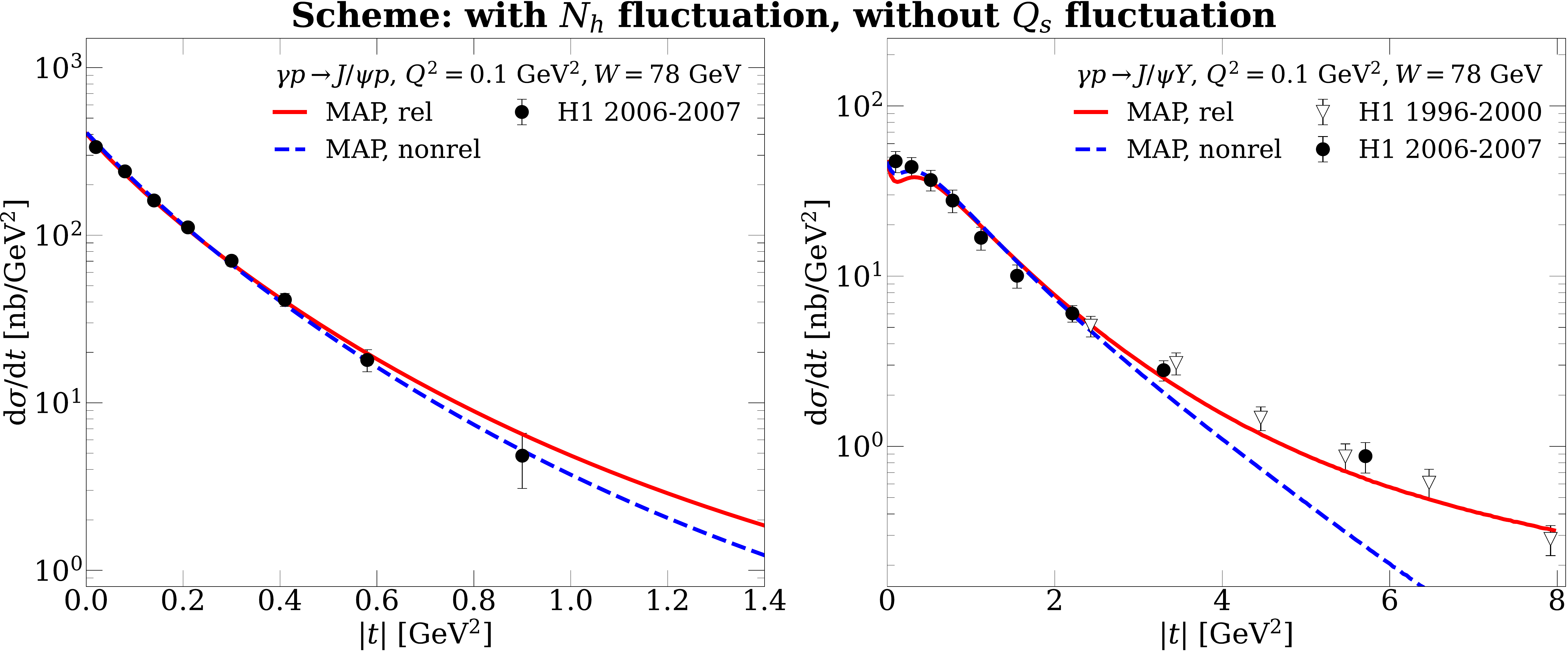}
    \caption{
    Fit result compared to the H1 data~\cite{H1:2013okq,H1:2003ksk}.
    }
    \label{fig:datacomparison_w_Nh_wo_Qs}
\end{figure*}

As an alternative scenario, we consider a model in which only fluctuations in the number of hot spots $N_h$ are included, while the fluctuations in the saturation scale $Q_s$ are absent. The results of this fit are presented in \cref{fig:corner_plot_w_Nh_wo_Qs,fig:datacomparison_w_Nh_wo_Qs,tab:params_w_Nh_wo_Qs}. 
Again, the hot spot size parameter $r_H$ peaks at zero in the non-relativistic limit, but otherwise all model parameters are rather well constrained by the HERA data. In particular, the mean number of hot spots $\overline{N_h}$ is constrained to be close to $3$ in both setups.
In this case, the MAP parameter sets yield reasonably good agreement with the H1 data. This agreement is particularly notable in the scenario that incorporates the relativistic correction. Nonetheless, we note that the description of the incoherent cross section at very small $t$ is worse than in the setup with $Q_s$ fluctuations described above. 

\begin{figure*}[ht!]
  \begin{minipage}[b]{.55\linewidth}
    \centering
    \includegraphics[width=\linewidth]{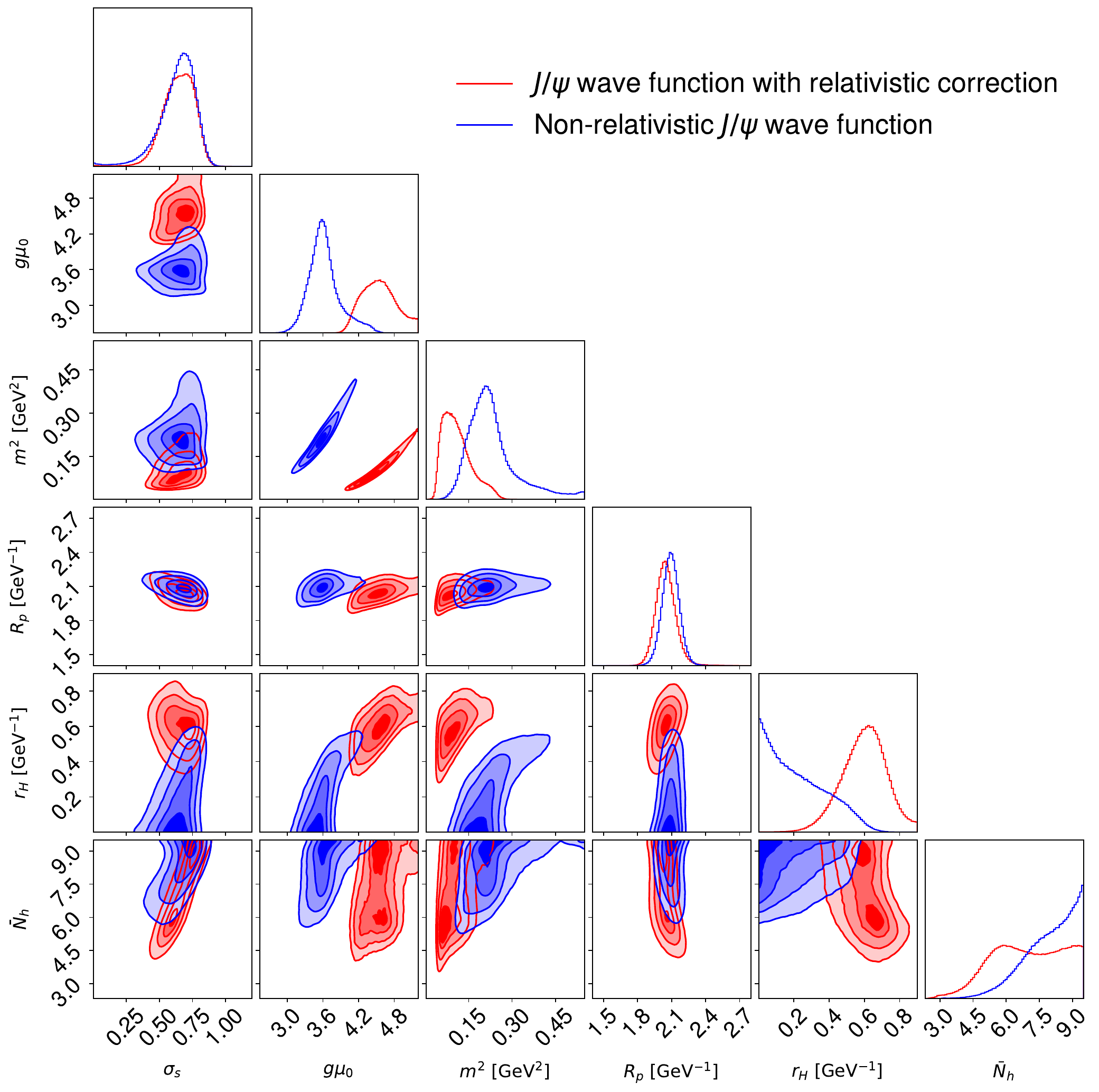}
    \captionof{figure}{Posterior distribution of parameters in the scheme with $Q_s$ fluctuation and with $N_h$ fluctuation.}%
    \label{fig:corner_plot_w_Nh_w_Qs}
  \end{minipage}\hspace{0.1em}
\begin{minipage}[b]{.42\linewidth}
    \centering
    {\renewcommand{\arraystretch}{2}
    \setlength{\tabcolsep}{0.2em}
    \begin{tabular}{ *{1}{c}*{1}{c}*{1}{c}*{1}{c} }
    \hline
      {Parameters} & {Prior range} & {MAP~(R)} & {MAP~(NR)}  \\
      \hline
      $\sigma_s$ & $[0.0, 1.2]$  & $\cpm{0.571}{0.145}{0.281}$ & $\cpm{0.685}{0.321}{0.184}$\\
      $g\mu_0$ & $[2.5, 5.2]$  & $\cpm{4.306}{0.249}{0.821}$ & $\cpm{3.467}{0.326}{0.783}$\\
      $m^2~[\rm GeV^2]$ & $[0.002,0.55]$ & $\cpm{0.069}{0.033}{0.147}$& $\cpm{0.175}{0.084}{0.258}$\\
      $R_p~[\rm GeV^{-1}]$ & $[1.4,2.8]$ & $\cpm{2.048}{0.167}{0.176}$ & $\cpm{2.057}{0.115}{0.183}$\\
      $r_H~[\rm GeV^{-1}]$ & $[0.0, 0.9]$ & $\cpm{0.578}{0.231}{0.264}$ & $\cpm{0.0002}{0.0002}{0.5278}$\\
      ${\overline{N}_h}$ & $[0.5, 9.5]$ & $\cpm{5.585}{1.164}{3.915}$ & $\cpm{9.012}{3.186}{0.488}$\\
      \hline
     \multicolumn{2}{c}{$\chi^2/{\rm d.o.f}$} & $0.45$ & $1.08$\\
      \hline
    \end{tabular}
    \vspace{2.7em}
    \captionof{table}{Parameter estimation from the posterior distribution in \cref{fig:corner_plot_w_Nh_w_Qs}.  }
    \label{tab:params_w_Nh_w_Qs}
    }
  \end{minipage}
\end{figure*}

\begin{figure*}[ht!]
    \centering
    \includegraphics[width=\textwidth]{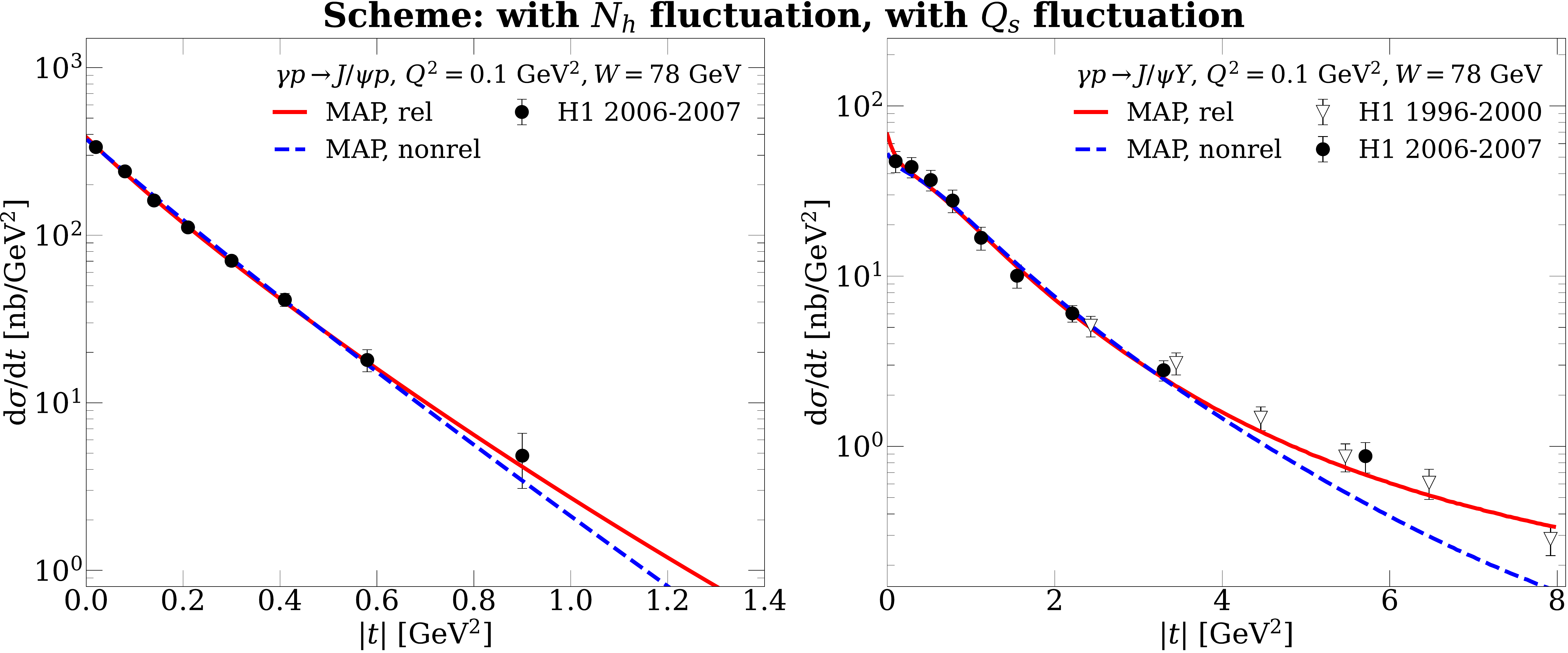}
    \caption{
    Fit result compared to the H1 data~\cite{H1:2013okq,H1:2003ksk}.
    }
    \label{fig:datacomparison_w_Nh_w_Qs}
\end{figure*}

In the most comprehensive scenario, the fluctuations in both $N_h$ and $Q_s$ are incorporated into the analysis. The outcomes of this scenario are presented in \cref{fig:corner_plot_w_Nh_w_Qs,fig:datacomparison_w_Nh_w_Qs,tab:params_w_Nh_w_Qs}. When the relativistic correction to the \jpsi  wave function is included, the analysis reveals that the MAP parameter set achieves  a remarkably good agreement with the experimental data. 
In this setup, we find that the H1 data do not constrain the mean number of hot spots, which is strongly correlated with the magnitude of the $Q_s$ fluctuations, as discussed above in the case with $Q_s$ fluctuations and a fixed number of hot spots.
In the fully non-relativistic limit, the MAP parameter set provides a reasonably good description of the experimental data at $t<4 ~\gev^2$. This outcome suggests that the model retains substantial predictive power even in the absence of relativistic corrections, particularly at small enough $|t|$.  

\section{Conclusion}
\label{sec:conclusion}

In this work, we have investigated exclusive \jpsi production in the dilute limit with an emphasis on a more comprehensive treatment of event-by-event fluctuations in the hotspot substructure of the target proton. Beyond the previously considered geometry and color charge fluctuations, we incorporated additional sources of fluctuations, allowing the number of hot spots $N_h$ and the saturation scale $Q_s$ associated with each hot spot to fluctuate from event to event. Furthermore, we included the leading relativistic correction to the 
\jpsi wave function, which provides a more accurate representation of the heavy quarkonium structure, and quantified its impact on the production cross section. The formalism developed here enabled us to derive (semi-)analytical results for both coherent and incoherent production, offering both theoretical clarity and computational efficiency. The obtained cross sections are \cref{eq:coherent_xsec_final,eq:total_xsection_final}.

Model parameters were constrained through a Bayesian analysis, ensuring a statistically robust determination of the proton substructure properties and associated uncertainties. 
The obtained posterior distributions are available in the Supplementary Material.
The results demonstrated that including additional sources of fluctuations beyond the hotspot geometry, together with relativistic corrections, significantly improves the agreement with experimental data and provides valuable insights into the fluctuating nature of the proton substructure. In particular, both the $N_h$ and $Q_s$ fluctuations improve the quality of the fit. The latter is found to be more important, especially at small $|t|$. Furthermore, we demonstrated that even if no $Q_s$ or $N_h$ fluctuations are included, it is still possible to get a reasonably good description of the HERA data when relativistic corrections are included. 
 Notably, the leading relativistic correction is also essential for describing the transition of the incoherent cross section to a power-law behavior at $|t|\gtrsim 3.5~\gev^2$.

\begin{acknowledgments}
This work was supported by the Research Council of Finland, the Centre of Excellence in Quark Matter (project 346324 and 364191) and projects 338263, 346567 and 359902, and by the European Research Council (ERC, grant agreements ERC-2023-101123801 GlueSatLight and ERC-2018-ADG-835105 YoctoLHC).
The content of this article does not reflect the official opinion of the European Union and responsibility for the information and views expressed therein lies entirely with the authors.

\end{acknowledgments}

\appendix
\section{The $c\bar c$ wave functions of $\gamma^*$}
\label{sec:appendix_gamma_wf}

The wave functions of $\gamma^*$ to a charm-anticharm ($c\Bar{c}$) dipole can be calculated using light-front perturbation theory. They can be found in the literature, e.g. Refs.~\cite{Brodsky:1997de,Kovchegov_Levin_2012,Lappi:2020ufv,Angelopoulou:2023qdm}. For longitudinal polarization of the photon ($\lambda=0$), the wave function reads 

\begin{equation}
    \label{eq:photon_wf_0}
    \begin{aligned}
        \Psi^{\lambda=0}_{\gamma^*;hh'} &= - e_q e \sqrt{N_c} \delta_{h,-h'} 2Qz(1-z) \frac{K_0(\varepsilon_z |\rt|)}{2\pi},
    \end{aligned}
    \end{equation}
while for the transverse polarizations ($\lambda=\pm1$), the corresponding wave functions are given by
\begin{subequations}
    \begin{equation}
    \label{eq:photon_wf_p1}
    \begin{aligned}
        &\Psi^{\lambda=+1}_{\gamma^*;hh'} = - e_q e \sqrt{2N_c} \left[ m_c\frac{K_0(\varepsilon_z |\rt|)}{2\pi} \delta_{h+}\delta_{h'+} \right.\\
        &\left. + \left(z\delta_{h+}\delta_{h'-} - (1-z)\delta_{h-}\delta_{h'+}\right) ie^{i\theta_r}  \frac{\varepsilon K_1(\varepsilon_z |\rt|)}{2\pi} \right] ,
    \end{aligned}
    \end{equation}
    \begin{equation}
    \label{eq:photon_wf_m1}
    \begin{aligned}
        &\Psi^{\lambda=-1}_{\gamma^*;hh'} = - e_q e \sqrt{2N_c} \left[ m_c\frac{K_0(\varepsilon_z |\rt|)}{2\pi} \delta_{h-}\delta_{h'-} \right.\\
        &\left. + \left((1-z)\delta_{h+}\delta_{h'-} - z\delta_{h-}\delta_{h'+}\right) ie^{-i\theta_r}  \frac{\varepsilon K_1(\varepsilon_z |\rt|)}{2\pi} \right]. 
    \end{aligned}
    \end{equation}
\end{subequations}
Here $h$ and $h'$ are helicities of $c$ and $c'$, respectively, and $\epsilon_z = z(1-z)Q^2 + m_c^2$. 

\section{The $c\bar c$ wave functions of \jpsi to the first relativistic correction }
\label{sec:appendix_Jpsi_wf}
The leading-order light front wave function of heavy quarkonia with the first relativistic correction can be computed based on the long-distance matrix elements constrained by decay width analyses in the non-relativistic QCD  (see Ref.~\cite{Lappi:2020ufv}).  In the following, we list the wave functions for different polarizations and helicities for \jpsi obtained in Ref~\cite{Lappi:2020ufv}.  
\begin{subequations}
    \begin{equation}
     \label{eq:jpsi_wf_0}
     \begin{aligned}
         &\Psi^{\lambda=0}_{\mathrm{J}/\psi;+-} = \Psi^{\lambda=0}_{\mathrm{J}/\psi;-+} = \frac{\pi\sqrt{2}}{\sqrt{m_c}}   \left\{ A \delta\left(z-\frac{1}{2}\right) \right.\\
         &\left. + \frac{B}{m_c^2}\left[ \left(\frac{5}{2} + r^2m_c^2\right)\delta\left(z-\frac{1}{2}\right)  - \frac{1}{4}\partial^2_z \delta\left(z-\frac{1}{2}\right)\right] \right\} ,
     \end{aligned}
    \end{equation}
    \begin{equation}
     \label{eq:jpsi_wf_pp_mm}
     \begin{aligned}
         &\Psi^{\lambda=1}_{\mathrm{J}/\psi;++} = \Psi^{\lambda=-1}_{\mathrm{J}/\psi;--} = \frac{2\pi}{\sqrt{m_c}}   \left\{ A \delta\left(z-\frac{1}{2}\right) \right.\\
         &\left. + \frac{B}{m_c^2}\left[ \left(\frac{7}{2} + r^2m_c^2\right)\delta\left(z-\frac{1}{2}\right)  - \frac{1}{4}\partial^2_z \delta\left(z-\frac{1}{2}\right)\right] \right\},
     \end{aligned}
    \end{equation}
    \begin{equation}
     \label{eq:jpsi_wf_pm_mp}
     \begin{aligned}
         &\Psi^{\lambda=1}_{\mathrm{J}/\psi;+-} = -\Psi^{\lambda=1}_{\mathrm{J}/\psi;-+} = (\Psi^{\lambda=-1}_{\mathrm{J}/\psi;-+})^* = (-\Psi^{\lambda=-1}_{\mathrm{J}/\psi;+-})^* \\
         &= -\frac{2\pi i}{m_c^{3/2}}B \delta\left(z-\frac{1}{2}\right) (r_1 + ir_2) , 
     \end{aligned}
    \end{equation}
    \begin{equation}
    \label{eq:eq:jpsi_wf_zero_components}
            \Psi^{\lambda=1}_{\mathrm{J}/\psi;--} = \Psi^{\lambda=-1}_{\mathrm{J}/\psi;++} = \Psi^{\lambda=0}_{\mathrm{J}/\psi;--} = \Psi^{\lambda=0}_{\mathrm{J}/\psi;++} = 0.
    \end{equation}
\end{subequations}

\bibliographystyle{JHEP-2modlong}
\bibliography{refs}

\end{document}